\documentclass[prb,twocolumn,amsmath,amssymb]{revtex4}

\usepackage{graphicx}
\usepackage{bm}

\begin{document}
\title{Entropy in Nonequilibrium Statistical Mechanics}

\author{Takafumi Kita}
\affiliation{Department of Physics, Hokkaido University, Sapporo 060-0810, Japan}

\date{\today}

\begin{abstract}
Entropy in nonequilibrium statistical mechanics is investigated theoretically
so as to extend the well-established equilibrium framework
to open nonequilibrium systems.
We first derive a microscopic expression of nonequilibrium
entropy for an assembly of identical bosons/fermions interacting via
a two-body potential.
This is performed by starting from the Dyson equation on the Keldysh contour
and following closely the procedure of Ivanov, Knoll and Voskresensky [Nucl.\ Phys.\ A
{\bf 672} (2000) 313].
The obtained expression is identical in form with an exact expression 
of equilibrium entropy and
obeys an equation of motion which satisfies the $H$-theorem
in a limiting case.
Thus, entropy can be defined unambiguously in nonequilibrium systems
so as to embrace equilibrium statistical mechanics.
This expression, however, differs from the one obtained by Ivanov {\em et al}., 
and we show explicitly that their ``memory corrections'' are not
necessary.
Based on our expression of nonequilibrium entropy, 
we then propose the following principle of 
maximum entropy for nonequilibrium steady states: 
``The state which is realized most probably 
among possible steady states without time evolution
is the one that makes entropy maximum as a function
of mechanical variables,
such as the total particle number, energy, momentum, energy flux, etc.''
During the course of the study, we also 
develop a compact real-time perturbation expansion 
in terms of the matrix Keldysh Green's function.
\end{abstract}

\maketitle

\section{Introduction}

Much effort has been directed towards extending equilibrium thermodynamics and
statistical mechanics to open nonequilibrium systems with flows of 
particles, momentum and/or energy.\cite{GP71,Haken75,Kuramoto84,CH93}
Beyond the linear-response theory,\cite{Kubo57,Zubarev74}
however, there seems to have been yet no established theoretical 
framework comparable to the equilibrium one.\cite{Gibbs,LL80}
The purpose of the present paper is to make a contribution to this 
fundamental issue,
especially on steady states without time evolution, 
by studying the roles of entropy in nonequilibrium statistical mechanics.

The present approach with entropy is motivated by
a couple of following observations.
First, equilibrium statistical mechanics is constructed on the principle
of equal {\em a priori} probabilities, and 
the equilibrium is identified as the one which is most probable.
It is hard to imagine that this concept of ``maximum probability'' 
loses its validity as soon as the system is driven from outside.
For example, a gas initially prepared in one half of the container
is expected to expand over the whole available space even in the
presence of heat conduction through the surface.
In equilibrium, it is entropy that embodies ``maximum probability,''
whose statistical mechanical expression is given by 
Boltzmann's principle:\cite{Cercignani98}
\begin{equation}
S_{\rm eq}\!=\! k_{\rm B}\log W \, .
\label{S-B1}
\end{equation}
And all the other free energies stem from $S_{\rm eq}$ through
the mathematical procedure
of Legendre transformations, thereby
inheriting the extremum property of entropy.
Thus, one may expect to have an appropriate description of
open steady states by extending the concept of entropy
or ``maximum probability''
to nonequilibrium situations.
Note in this context that eq.\ (\ref{S-B1}), which represents the principle of
equal {\em a priori} probabilities,
is essentially an equilibrium
expression with no dynamical equation attached to it.

The second motivation originates from the classical 
Boltzmann equation.\cite{CC90,HCB54,Cercignani88,Cercignani98}
Let us define the total entropy of dilute gases by
\begin{equation}
S=-k_{\rm B}\int\frac{{\rm d}^{3}r{\rm d}^{3}p}{(2\pi\hbar)^{3}}\, f(\log f-1) \, ,
\label{S-B2}
\end{equation}
where $f\!=\! f({\bm p},{\bm r},t)$ denotes the distribution function.\cite{comment1}
Following the procedure to prove the $H$-theorem,\cite{Cercignani88}
we then obtain the inequality:
\begin{equation}
\frac{{\rm d}S}{{\rm d}t}
+\int{\bm\nabla}\cdot {\bm j}_{S}({\bm r})\,{\rm d}^{3}r \geq 0\, ,
\label{dS/dt>0}
\end{equation}
with ${\bm j}_{S}({\bm r})\!\equiv\! 
-k_{\rm B}\int\frac{{\rm d}^{3}p}{(2\pi\hbar)^{3}} 
\frac{{\bm p}}{m} f(\log f\!-\!1)$.
It hence follows that ${\rm d}S/{\rm d}t\!\geq\! 0$
for the isolated system.
This is the usual $H$-theorem.
Looking at eq.\ (\ref{dS/dt>0}) more carefully, however,
one may notice that ${\rm d}S/{\rm d}t\!\geq\! 0$ holds 
whenever $\int{\bm\nabla}\!\cdot\! {\bm j}_{S}({\bm r}){\rm d}^{3}r\!=\!0$.
Thus, entropy is expected to increase monotonically even in open systems
as long as there is no net inflow or outflow of entropy through the boundary, 
besides those of energy, 
momentum and particles.
This observation suggests that the open steady state, if it exists,
may also correspond to the maximum of entropy with appropriately chosen 
independent variables.
In this context, eq.\ (\ref{S-B2}) is superior to eq.\ (\ref{S-B1}) 
in that it is applicable to nonequilibrium systems,
but inferior to eq.\ (\ref{S-B1}) in that it is good only for 
dilute classical gases.

Now, the main purposes of the present paper are twofold.
First, we derive an expression of nonequilibrium
entropy for an assembly of identical bosons/fermions
interacting via a two-body potential
so as to be compatible with equilibrium statistical mechanics.
Such an investigation was performed recently for the contact interaction 
in a seminal paper by Ivanov, Knoll and Voskresensky.\cite{Ivanov99,Ivanov00} 
We here extend their consideration to general two-body interactions,
critically reexamine their derivation,\cite{Ivanov00}
and present an expression of nonequilibrium
entropy which differs from theirs
in an essential point.
To be more specific, we adopt the nonequilibrium Dyson equation for the Keldysh
matrix\cite{Keldysh64} as our starting point, which is transformed into
a tractable form by the gradient expansion, i.e.,
the procedure well-known for a microscopic derivation of quantum transport 
equations.\cite{KB62,Keldysh64,Rainer83,RS86,BM90,LF94,HJ98,Ivanov99,Ivanov00,Kita01}
An expression of nonequilibrium entropy density is then obtained 
from the reduced Dyson equation as eq.\ (\ref{entropy-density}) below.
It is a direct extension of eq.\ (\ref{S-B2}) to include
both quantum and many-body effects in nonequilibrium
situations, which is also compatible with the equilibrium expression.\cite{Kita99}
We will show explicitly that ``memory corrections'' of 
Ivanov {\em et al}.,\cite{Ivanov00} which is the origin of the above mentioned difference
in nonequilibrium entropy, 
are both unnecessary and incompatible with equilibrium statistical mechanics.

Second, we propose a principle of maximum entropy for nonequilibrium
steady states in \S\ref{subsec:Smax}.
A key point is that we choose mechanical variables,
such as the total particle number, energy, momentum, energy flux, etc.,
as independent variables of entropy.
Indeed, temperature, pressure and chemical potential,
which are not adopted here, are all
equilibrium thermodynamic variables 
defined with partial derivatives of eq.\ (\ref{S-B1});
thus, they cannot specify any nonequilibrium state of the system.
The principle may enable microscopic treatments of 
open steady states in exactly the same way as equilibrium systems.
Its validity can only be checked by
its consistency with experiments, as is the case for the principle of
equal {\em a priori} probabilities in equilibrium statistical mechanics.
Thus, it will be tested in the next paper on
Rayleigh-B\'enard convection\cite{CH93,Chandrasekhar61,Busse78,Croquette89,
Koschmieder93,deBruyn96,BPA00}
of a dilute classical gas, which may be regarded as the canonical system of
nonequilibrium steady states with pattern formation.
It may be worth emphasizing at this stage that the present principle is connected 
with entropy itself. 
Thus, it has to be distinguished from the principle of excess entropy production
by Gransdorff and Prigogine,\cite{GP71} which has been criticized by
Graham,\cite{Graham81} for example;
see also ref.\ \onlinecite{KP98}.

During the course of study, 
we also develop a compact perturbation expansion
on the Keldysh contour.\cite{Keldysh64}
In principle, this expansion can be carried out 
for the round-trip Keldysh contour
in the same way as in the equilibrium theory.\cite{LW60,AGD,FW}
When writing it with respect to 
the real-time contour of $-\infty\!\leq\! t\!\leq\!\infty$,
however, one usually has to introduce additional contour indices\cite{Keldysh64,RS86}
which make the actual calculations rather cumbersome
and complicated.
The present method will enable us to carry out the expansion
on the real-time contour directly
in terms of the $2\times 2$ Keldysh Green's function
without using the contour indices.
Among various approximations in the perturbation expansion, 
we here specifically consider Baym's $\Phi$-derivative approximation.\cite{Baym62}
It has at least the following advantages: (i) it includes the exact theory;
(ii) various conservation laws are automatically obeyed;
(iii) the vertex corrections, or the Landau Fermi liquid corrections
in a different terminology, are naturally included;
(iv) $n$-particle ($n\!=\! 2,3,\cdots$) correlations can also be calculated
within the same approximation scheme, i.e.,
there is a definite prescription here 
to treat the Bogoliubov-Born-Green-Kirkwood-Yvons (BBGKY) hierarchy.\cite{Cercignani88}
The derivation of nonequilibrium entropy by Ivanov {\em et al}.\ \cite{Ivanov00}
will be reexamined critically within the present expansion scheme.

This paper is organized as follows.
In \S 2, we develop a compact real-time perturbation expansion 
in terms of the matrix Keldysh Green's function
for an assembly of identical 
bosons/fermions interacting via a two-body potential.
We consider the $\Phi$-derivative approximation in detail to
write down the Dyson equation for the Green's function 
and the expression for the two-particle correlation function.
In \S 3, we first introduce the spectral function $A$ and 
the distribution function $\phi$ 
in the Wigner representation; they form alternative two independent 
components of the Keldysh Green's function.
We then carry out the first-order gradient expansion to the Dyson equation
to obtain the equations for $A$ and $\phi$.
In \S 4, we derive an expression of nonequilibrium entropy 
as eq.\ (\ref{entropy-density}) below.
A detailed discussion will be given on the difference between the present
expression and the one obtained by 
Ivanov {\em et al}.\cite{Ivanov00}
We then propose in \S \ref{subsec:Smax} 
a principle of maximum entropy for nonequilibrium steady states.
Section 5 summarizes the paper.
In Appendix\ref{App:conserve}, we show
with the present perturbation-expansion scheme 
that various conservation laws are
automatically obeyed in the $\Phi$-derivative approximation.
Appendix\ref{App:Gamma} presents expressions of the vertex functions
in the second-order $\Phi$-derivative approximation.
In Appendix\ref{app:conserve-W}, 
we derive basic conservation laws
in the first-order gradient expansion of the $\Phi$-derivative
approximation.
Finally in Appendix\ref{app:Phi}, we identify 
the origin of the difference on equilibrium entropy
between refs.\ \onlinecite{Kita99} and \onlinecite{CP75}
to confirm that the ``memory corrections''
by Ivanov {\em et al}.\ \cite{Ivanov00} are
unnecessary.

\section{\label{sec:perturb}Perturbation expansion with Keldysh matrix}

\subsection{Contour-ordered Green's function}

We consider an assembly of identical bosons/fermions
whose total Hamiltonian at time $t$ is given by
\begin{equation}
{\cal H}(t) = H_{0} +H'(t)+H_{\rm int}a(t)  \, .
\label{calH}
\end{equation}
Here $H_{0}$ denotes the kinetic energy,
$H'(t)$ is a one-body time-dependent perturbation
satisfying $H'(-\infty)\!=\! 0$,
$H_{\rm int}$ is a two-body interaction, 
and $a(t)$ is some adiabatic factor given
by $a(t)\!=\!\theta(-t){\rm e}^{0_{+}t}\!+\!\theta(t)$, for example, with
$\theta$ the step function and $0_{+}$ an infinitesimal positive constant.
The system at $t\!=\!-\infty$ is assumed to be in some
thermodynamic state described by a density matrix
corresponding to $H_{0}$.
Thus, we here need not consider from the beginning the 
contribution from the path along the
imaginary time axis,\cite{RS86,HJ98} i.e., ``initial correlation,''
in the perturbation expansion with
respect to $H_{\rm int}$.

The explicit expressions of $H_{0}$, $H'(t)$ and $H_{\rm int}$ 
are given in second quantization by
\begin{subequations}
\label{H_0int}
\begin{equation}
H_{0}=\int\!
\psi^{\dagger}({\bm r})\frac{-\hbar^{2}}{2m}\nabla^{2}
\psi({\bm r}) \, {\rm d}^{3}r\, ,
\label{H_0}
\end{equation}
\begin{equation}
H'(t)=\int\!
\psi^{\dagger}({\bm r})U({\bm r},t)
\psi({\bm r}) \,{\rm d}^{3}r\, ,
\label{H'}
\end{equation}
\begin{equation}
H_{\rm int}
=\frac{1}{2}
\int\!{\rm d}^{3}r\!\int\!{\rm d}^{3}r'
V({\bm r}\!-\!{\bm r}')\psi^{\dagger}({\bm r})
\psi^{\dagger}({\bm r}')
\psi({\bm r}')\psi({\bm r}) 
\, .
\label{H_int}
\end{equation}
\end{subequations}
Here $m$ is the particle mass, $U$ is an external potential, 
and $V({\bm r})\!=\!V(|{\bm r}|)$ is a two-body interaction 
which can be expanded in Fourier series as
\begin{equation}
V({\bm r})=\int\! \frac{{\rm d}^{3}q}{(2\pi \hbar)^{3}}\,
V_{{\bm q}}{\rm e}^{i{\bm q}
\cdot{\bm r}/\hbar} \, .
\label{V-Fourier}
\end{equation}
The spin degrees of freedom will be 
suppressed for the time being.

We now adopt the interaction representation 
with respect to $H^{(0)}(t)\!\equiv\! H_{0} \!+\!H'(t)$.
Then, the time evolution of the system is
described by the unitary operator:
\begin{equation}
S_{C}\equiv T_{C}\exp\left[-\frac{i}{\hbar}\int_{C}H_{\rm int}(t^{C}) 
{\rm d}t^{C} \right] .
\label{S_C}
\end{equation}
Here $C$ is a round-trip contour along the real-time axis
from $t\!=\!-\infty$ towards $t\!=\!\infty$,\cite{RS86,HJ98}
$T_{C}$ denotes the contour-ordering operator along $C$,
and $H_{\rm int}(t^{C})$ is the interaction representation 
of $H_{\rm int}a(t)$ with respect to $H^{(0)}(t)$.\cite{RS86,HJ98}

We next introduce the contour-ordered Green's function by
\begin{equation}
G(1^{C},2^{C})
=-\frac{i}{\hbar}
\langle T_{C}\psi_{{\cal H}}(1^{C})\psi_{{\cal H}}^{\dagger}(2^{C})\rangle
\, , 
\label{G-def}
\end{equation}
where $\psi_{{\cal H}}(1^{C})$ is the Heisenberg operator for $\psi({\bm r}_{1})$
with $1^{C}\!\equiv\!{\bm r}_{1}t_{1}^{C}$.
The perturbation expansion of $G$ with respect to $H_{\rm int}$
may be carried out in the same way as in
the equilibrium theory\cite{LW60,AGD,FW}
by using Feynman diagrams.
Indeed, one only needs to change the imaginary-time contour
of the equilibrium theory into the real-time contour $C$.
It hence follows that $G$ satisfies the Dyson equation:\cite{LW60}
\begin{eqnarray}
&&\hspace{-10mm}
\left[i\hbar \frac{\partial }{\partial t_{1}^{C}}\!+\! 
\frac{\hbar^{2}\nabla_{1}^{2}}{2m}\!-\!U(1^{C})
\right]\!
G(1^{C},2^{C})
\nonumber \\
&&\hspace{-10mm}
-
\int \! \Sigma(1^{C},3^{C})
G(3^{C},2^{C})\,{\rm d}3^{C} =\delta(1^{C},2^{C}) \, ,
\label{Dyson}
\end{eqnarray}
where $\Sigma$ denotes the irreducible self-energy.
However, the round-trip contour $C$ is not convenient
for practical calculations, since time $t$ appears twice 
on $C$ with different orders.
Thus, we usually have to introduce additional contour indices
to distinguish them,\cite{Keldysh64,RS86}
which make the actual calculations rather cumbersome
and complicated.

\subsection{Feynman rules in Keldysh space}

It is desirable to find a simple and compact method 
to carry out the perturbation expansion directly 
on the real-time contour of $-\infty\!\leq\! t\!\leq\!\infty$.
This is possible for the two-body interaction 
of eq.\ (\ref{H_int}) (and also for impurity potentials), 
as explained below.
We first divide the contour $C$
into $C_{1}$ and $C_{2}$, each running from $-\infty$ to $\infty$
and from $\infty$ to $-\infty$, respectively.
Accordingly, we write the integration in eq.\ (\ref{S_C}) as a sum of the two
contributions:
\begin{equation}
\int_{C} {\rm d}t^{C}=\int_{-\infty(C_{1})}^{\infty}{\rm d}t
-\int_{-\infty(C_{2})}^{\infty}{\rm d}t \, .
\label{int_C}
\end{equation}
We next introduce the vector:
\begin{equation}
\vec{\psi}(1)=\left[\begin{array}{c}
\vspace{1mm}
\psi(1^{1}) \\ \psi(1^{2})
\end{array}\right] ,\hspace{5mm}
\vec{\psi}^{\dagger}(1)\equiv\left[
\psi^{\dagger}(1^{1})\,\,\psi^{\dagger}(1^{2})\right] ,
\label{vec-psi}
\end{equation}
where $\psi(1^{j})$ 
denotes the interaction representation of $\psi({\bm r}_{1})$
with $1^{j}\!\equiv\! {\bm r}_{1}t_{1}^{C_{j}}$.
Then eq.\ (\ref{S_C}) can be rewritten in terms of
$\vec{\psi}^{\dagger}$, $\vec{\psi}$ 
and the normal-ordering operator\cite{FW} ${\cal N}$ as 
\begin{eqnarray}
&&\hspace{-10mm}
S_{C}=T_{C}\exp\!\biggl[-\frac{i}{\hbar}\int{\rm d}1 \int{\rm d}1'\,
\frac{{\cal N}}{2}
\bar{V}(1\!-\!1')
\nonumber \\
&&\hspace{10mm}\times
\vec{\psi}^{\dagger}(1)\vec{\psi}(1)
\vec{\psi}^{\dagger}(1')\check{\tau}_{3}
\vec{\psi}(1')\biggr] ,
\label{S_C2}
\end{eqnarray}
where $1\!\equiv\!{\bm r}_{1}t_{1}$ with $-\infty\!\leq\!t_{1}\!\leq\!\infty$,
$\bar{V}$ is defined by $\bar{V}(1-1')\!\equiv\! \delta(t_{1}-t_{1}')
V({\bm r}_{1}-{\bm r}_{1}')$,
and $\check{\tau}_{3}$ is the third Pauli matrix.
The equivalence of eqs.\ (\ref{S_C}) and (\ref{S_C2}) may be checked
easily by writing eq.\ (\ref{S_C2}) without using ${\cal N}$.
The interaction in eq.\ (\ref{S_C2}) can be expressed diagrammatically as Fig.\ 1.
The expression (\ref{S_C2}) is quite useful for our purpose,
because (i) the pairs $\vec{\psi}^{\dagger}(1)\vec{\psi}(1)$ and
$\vec{\psi}^{\dagger}(1')\check{\tau}_{3}
\vec{\psi}(1')$ can be moved around anywhere within the ${\cal N}$ and/or $T_{C}$
operators in the perturbation expansion, 
and (ii) a contraction of $\vec{\psi}(i)$ 
with $\vec{\psi}^{\dagger}(j)$ 
automatically yields a $2\!\times\!2$ matrix 
$\langle T_{C}\vec{\psi}(i)\vec{\psi}^{\dagger}(j)\rangle_{0}$,
where the subscript $0$ denotes the average with respect to $H^{(0)}$.
Also, the final contraction within a closed particle loop can be transformed 
as $\langle T_{C} \vec{\psi}^{\dagger}(i)\check{\cal M}(i,j)\vec{\psi}(j)\rangle_{0}
\!=\!{\rm Tr}\check{\cal M}(i,j)\langle T_{C} \vec{\psi}(j)
\vec{\psi}^{\dagger}(i)\rangle_{0}$ with $\check{\cal M}(i,j)$ denoting
some matrix product of contractions.

\begin{figure}[t]
\begin{center}
  \includegraphics[width=0.4\linewidth]{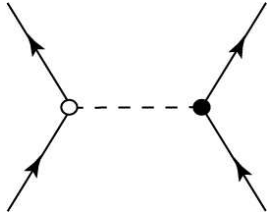}
\end{center}
  \caption{A diagrammatic expression of the interaction in eq.\ (\ref{S_C2}).
  The outgoing and incoming lines represent $\vec{\psi}^{\dagger}$ and $\vec{\psi}$,
  respectively, whereas the broken line corresponds to the interaction potential.
  The open and filled circles denote the unit matrix and 
  the third Pauli matrix $\check{\tau}_{3}$,
  respectively.}
  \label{fig:1}
\end{figure}

Now, one may realize that the perturbation expansion can be 
carried out compactly on the real-time axis
without the ambiguity on the limits of time integrations\cite{KB62}
nor the complexity from the contour indices.\cite{Keldysh64,RS86}
We introduce the matrix Green's function for this purpose:
\begin{eqnarray}
&&\hspace{-10mm}
\check{G}(1,1')\equiv -\frac{i}{\hbar}
\langle T_{C}\vec{\psi}_{\cal H}(1)\vec{\psi}_{\cal H}^{\dagger}(1')
\rangle 
\nonumber \\
&&\hspace{1.5mm}=
\left[\begin{array}{cc}
G_{11}(1,1') & G_{12}(1,1')
\\
G_{21}(1,1') & G_{22}(1,1')
\end{array}
\right] .
\label{check-G}
\end{eqnarray}
Here $G_{12}(1,1')
\!=\!\mp(i/\hbar)\langle\psi_{\cal H}^{\dagger}(1')\psi_{\cal H}(1)\rangle
\!\equiv\! G^{<}(1,1')$ 
and $G_{21}(1,1')
\!=\!-(i/\hbar)\langle\psi_{\cal H}(1)\psi_{\cal H}^{\dagger}(1')\rangle
\!\equiv\! G^{>}(1,1')$ 
are the correlation functions introduced by Kadanoff and Baym\cite{KB62}
with the upper (lower) sign corresponding to bosons (fermions).
They satisfy
\begin{subequations}
\label{G-prop}
\begin{equation}
G_{12}(1,1')=
-G_{12}^{*}(1',1) \, ,\hspace{3mm}
G_{21}(1,1')=
-G_{21}^{*}(1',1) \, .
\label{G-prop1}
\end{equation}
The diagonal elements can be written explicitly with respect to 
the off-diagonal elements as
\begin{equation}
G_{11}(1,1')=\theta(t_{1}'\!-\!t_{1})G_{12}(1,1')+
\theta(t_{1}\!-\!t_{1}')G_{21}(1,1') \, ,
\label{G-prop2}
\end{equation}
\begin{equation}
G_{22}(1,1')=\theta(t_{1}\!-\!t_{1}')G_{12}(1,1')+
\theta(t_{1}'\!-\!t_{1})G_{21}(1,1') \, ,
\label{G-prop3}
\end{equation}
\end{subequations}
so that $G_{11}(1,1')\!=\!-G_{22}^{*}(1',1)$ and $G_{11}\!+\!G_{22}\! =\!
G_{12}\!+\!G_{21}$.
Thus, there are only two independent components in $\check{G}$,
i.e., $G_{12}$ and $G_{21}$.
However, all the four elements
are necessary in the perturbation expansion.
Equation (\ref{G-prop}) can be expressed compactly in terms of $\check{G}$ as
\begin{equation}
\check{G}(1,1')=-\check{\tau}_{1}\check{G}^{\dagger}(1',1)\check{\tau}_{1} \, ,
\hspace{5mm} {\rm Tr}\check{G}={\rm Tr}\check{G}\check{\tau}_{1} \, ,
\label{checkG-prop}
\end{equation}
with $\check{\tau}_{1}$ denoting the first Pauli matrix.
Equation (\ref{checkG-prop}) will be useful later.

The Feynman rules to calculate 
$\check{G}$ are summarized as follows:
(i) Draw all possible $n$th-order connected diagrams.
(ii) With each such diagram, associate a factor
\begin{equation}
\frac{(i\hbar)^{n}(\pm 1)^{\ell}}{2^{n}n! } \, ,
\label{Feynman-rule-G}
\end{equation}
where $\ell$ denotes the number of closed loops.
Note that topologically identical
diagrams appear $n!$ times.
(iii) For each line
arriving at $1$ from $2$,
associate the matrix
$\check{G}^{(0)}(1,2)$ or
$\check{\tau}_{3}\check{G}^{(0)}(1,2)$,
and multiply it from the left of the matrix arriving at
$2$.
(iv) If the time arguments of $\check{G}^{(0)}$ are equal,
we need the replacement $G_{11}^{(0)}$, $G_{22}^{(0)}\!\rightarrow\!G_{12}^{(0)}$
due to the operator ${\cal N}$ in eq.\ (\ref{S_C2}).
(v) Integrate and sum over all the internal 
variables, and take ${\rm Tr}$
for every closed particle line.
(vi) The spin degrees of freedom can be included easily
by multiplying every closed-loop contribution by $2S\!+\! 1$,
where $S$ denotes the magnitude of spin.

\begin{figure}[t]
\begin{center}
  \includegraphics[width=0.8\linewidth]{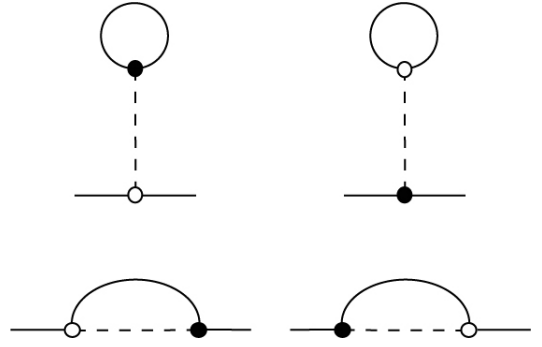}
\end{center}
  \caption{First-order diagrams for the matrix Green's function.}
  \label{fig:2}
\end{figure}

For example, Fig.\ 2 enumerates topologically distinct first-order diagrams
for $\check{G}$. 
The corresponding analytic expression 
is given by
\begin{eqnarray}
&&\hspace{-8.5mm}
\check{G}^{(1)}(1,1')
=
\frac{i\hbar}{2}\int {\rm d}2 \int {\rm d}2'\,
\bar{V}(2\!-\!2')\bigl\{
\pm\check{G}(1,2)
\nonumber \\
&&\hspace{1.5mm}
\times\bigl[\check{G}(2,1')
{\rm Tr}\check{\tau}_{3}\check{G}(2',2')+\check{\tau}_{3}\check{G}(2,1')
{\rm Tr}\check{G}(2',2')\bigr]
\nonumber \\
&&\hspace{1.5mm} 
+\check{G}(1,2)\bigl[\check{G}(2,2')
\check{\tau}_{3}+\check{\tau}_{3}\check{G}(2,2')\bigr]
\check{G}(2',1')\bigr\}
\, ,
\label{G^(1)}
\end{eqnarray}
where we have replaced $\check{G}^{(0)}$ by $\check{G}$ on the 
right-hand side to include the renormalization effects.

The Dyson equation (\ref{Dyson}) is transformed into
a matrix form as
\begin{eqnarray}
&&\hspace{-8mm}
\left[i\hbar \frac{\partial }{\partial t_{1}}\!+\! 
\frac{\hbar^{2}\nabla_{1}^{2}}{2m}\!-\!U(1)
\right]\!
\check{G}(1,1')
\nonumber \\
&&\hspace{-8mm}
-
\int \check{\Sigma}(1,2)\check{\tau}_{3}
\check{G}(2,1')\,{\rm d}2 
=\delta(1,1')\check{\tau}_{3} \, .
\label{Dyson-check}
\end{eqnarray}
The appearance of $\check{\tau}_{3}$ between $\check{\Sigma}$ and $\check{G}$
is due to eq.\ (\ref{int_C}),
whereas $\check{\tau}_{3}$ in front of $\delta(1,2)$ originates
from the anti-time ordering on $C_{2}$.
Equation (\ref{Dyson-check}) is expressed alternatively in an integral form as
\begin{eqnarray}
&&\hspace{-12mm}
\check{G}(1,1')
=\check{G}^{(0)}(1,1')
\nonumber \\
&&\hspace{-5mm}
+\int {\rm d}2\int  {\rm d}2'\, \check{G}^{(0)}(1,2)\check{\tau}_{3}
\check{\Sigma}(2,2')\check{\tau}_{3}
\check{G}(2',1') \, ,
\label{Dyson-check-2}
\end{eqnarray}
with
\begin{equation}
\check{G}^{(0)}(1,1')\equiv
\left[i\hbar \frac{\partial }{\partial t_{1}}
+\frac{\hbar^{2}\nabla_{1}^{2}}{2m}\!-\!U(1)\right]^{ -1}
\delta(1,1')\check{\tau}_{3} \, .
\end{equation}
Comparing eq.\ (\ref{G^(1)}) with the second term on the
right-hand side of eq.\ (\ref{Dyson-check-2}), 
we identify the first-order self-energy
with renormalization as
\begin{subequations}
\label{Sigma^(1)}
\begin{eqnarray}
&&\hspace{-5mm}
\check{\Sigma}^{(1)}(1,1')
\nonumber \\
&&\hspace{-5mm}
=\pm  \frac{i\hbar}{2}\delta(1,1')\! \int \!{\rm d}2\,
\bar{V}(1\!-\!2)\bigl[\, \check{1} \,
{\rm Tr}  \check{\tau}_{3}
\check{G}(2,2)
+ \check{\tau}_{3}{\rm Tr}  
\check{G}(2,2)
\bigr]
\nonumber \\
&&\hspace{-1.5mm} 
+
\frac{i\hbar}{2}
\bar{V}(1\!-\!1')
\bigl[\check{\tau}_{3}
\check{G}(1,1')
+\check{G}(1,1')\check{\tau}_{3}
\bigr] \, ,
\label{Sigma^(1)a}
\end{eqnarray}
where $\check{1}$ denotes the $2\!\times\! 2$ unit matrix.
One can check that the off-diagonal elements cancel
out in eq.\ (\ref{Sigma^(1)a}).
We also have to consider the Feynman rule (iv) above.
Equation (\ref{Sigma^(1)a}) is thereby simplified into
\begin{equation}
\check{\Sigma}^{(1)}(1,1')=\check{\tau}_{3}\Sigma^{{\rm HF}}(1,1')
\, ,
\label{Sigma^(1)b}
\end{equation}
\end{subequations}
where $\Sigma^{{\rm HF}}$ is the Hartree-Fock self-energy:
\begin{eqnarray}
&&\hspace{-10mm}
\Sigma^{{\rm HF}}(1,1')\equiv
\pm  \delta(1,1')\int{\rm d}2\,
\bar{V}(1\!-\!2)
i\hbar G_{12}(2,2)
\nonumber \\
&&\hspace{10mm}
+\bar{V}(1\!-\!1')i\hbar
G_{12}(1,1') \, .
\label{Sigma-HF}
\end{eqnarray}
Note $\Sigma^{{\rm HF}}(1,1')\!=\![\Sigma^{{\rm HF}}(1',1)]^{*}$,
i.e., it is Hermitian.

Thus, the above Feynman rules enable us 
a straightforward and automatic perturbation expansion of $\check{G}$
with neither
using the contour indices nor worrying about the ordering of $T_{C}$.

\subsection{$\Phi$-derivative approximation}

In carrying out practical calculations,
we are almost always obliged to introduce some kind of approximations.
In this context, Baym\cite{Baym62} presented an extremely useful approximation
scheme based on the skeleton expansion,\cite{LW60} i.e., 
the $\Phi$-derivative approximation.
The functional $\Phi\!=\!\Phi[G]$ 
was introduced by Luttinger and Ward as part of the
exact thermodynamic functional.\cite{LW60}
The $\Phi$-derivative approximation was successively suggested by 
Luttinger\cite{Luttinger60} in the equilibrium theory,
but has turned out to be especially useful for dynamical systems.\cite{Baym62}
It has the following advantages:
(i) it becomes exact if all the terms
in the skeleton expansion are retained;
(ii) various conservation laws, which have crucial importance to describe
dynamical systems, are obeyed automatically;
(iii) the vertex corrections, or the Landau Fermi liquid corrections
in a different terminology, are naturally included;
(iv) $n$-particle ($n\!=\! 2,3,\cdots$) correlations can be obtained 
with the same approximation scheme, i.e.,
there is a definite prescription here 
to treat the BBGKY hierarchy.\cite{Cercignani88}
A detailed study on the dynamical $\Phi$-derivative approximation 
has also been performed by 
Ivanov {\em et al}.\ \cite{Ivanov99,Ivanov00}
for the contact interaction.
We describe it for the general two-body interaction $V$
in terms of the present perturbation
expansion scheme.
It is shown explicitly in Appendix\ref{App:conserve} that various conservation laws 
are automatically satisfied in the $\Phi$-derivative approximation.

Let us define the functional $\Phi$ in terms of eq.\ (\ref{S_C2}) by
\begin{equation}
\Phi \equiv \bigl[\langle\ln S_{C}\rangle_{0} -1\bigr]_{{\rm skeleton},
\check{G}^{(0)}\rightarrow \check{G}} \, .
\label{Phi-def}
\end{equation}
Thus, $\Phi$ formally consists of infinite closed skeleton diagrams
with $\check{G}^{(0)}$ replaced by $\check{G}$.\cite{LW60}
The Feynman rules to calculate $\Phi$ are exactly
the same as those of $\check{G}$
which are given around eq.\ (\ref{Feynman-rule-G}).
The only care necessary is that we have
$(n\!-\!1)!$ topologically identical diagrams here.
The exact irreducible self-energy is obtained from $\Phi$ by
\begin{equation}
\check{\Sigma}(1,1')=\pm \check{\tau}_{3}
\frac{\delta \Phi}{\delta \check{G}(1',1)}
\check{\tau}_{3} \, .
\label{Sigma-Phi}
\end{equation}
The necessity of $\check{\tau}_{3}$ on both sides may be realized
from eq.\ (\ref{Dyson-check-2}).

The $\Phi$-derivative approximation denotes retaining 
some partial diagrams from the infinite series for $\Phi$
and determining $\check{G}$ and $\check{\Sigma}$ self-consistently
by eqs.\ (\ref{Dyson-check}) and (\ref{Sigma-Phi}).
It follows from eqs.\ (\ref{checkG-prop}) and (\ref{Sigma-Phi})
that $\check{\Sigma}$ thus obtained also satisfies
\begin{equation}
\check{\Sigma}(1,1')=-\check{\tau}_{1}\check{\Sigma}^{\dagger}(1',1)\check{\tau}_{1} \, .
\label{checkSigma-prop}
\end{equation}

\begin{figure}[t]
\begin{center}
  \includegraphics[width=0.8\linewidth]{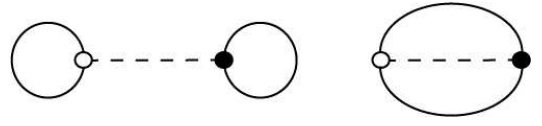}
\end{center}
  \caption{First-order diagrams for $\Phi$.}
  \label{fig:3}
\end{figure}

The first-order diagrams for $\Phi$ are given in Fig.\ 3.
They correspond to
\begin{eqnarray}
&&\hspace{-3mm}
\Phi^{(1)}=
\frac{i\hbar}{2}\int {\rm d}1 \int{\rm d}1'\, \bar{V}(1\!-\!1')
\left[{\rm Tr}\check{G}(1,1)
{\rm Tr}\check{\tau}_{3}\check{G}(1',1')
\right.
\nonumber \\
&&\hspace{8mm} \left.
\pm{\rm Tr}\check{G}(1,1')
\check{\tau}_{3}
\check{G}(1',1) \right] \, .
\label{Phi^(1)}
\end{eqnarray}
Then $\check{\Sigma}^{(1)}$
is calculated by eq.\ (\ref{Sigma-Phi})
to yield eq.\ (\ref{Sigma^(1)a}).

Next, Fig.\ 4 enumerates topologically distinct second-order diagrams.
The corresponding analytic expression is given by
\begin{eqnarray}
&&\hspace{-6mm}
\Phi^{(2)}=
\frac{(i\hbar)^{2}}{2^{2}2!}\int {\rm d}1 \int {\rm d}1' 
\int {\rm d}2\int {\rm d}2'\,
\bar{V}(1\!-\!1')\bar{V}(2\!-\!2')
\nonumber \\
&&\hspace{5mm}\times
\bigl[{\rm Tr}\check{G}(1,2)\check{G}(2,1)
{\rm Tr}\check{\tau}_{3}\check{G}(1',2')
\check{\tau}_{3}\check{G}(2',1')
\nonumber \\
&&\hspace{8mm}
+ {\rm Tr}\check{G}(1,2')
\check{\tau}_{3}\check{G}(2',1)
{\rm Tr}\check{\tau}_{3}\check{G}(1',2)
\check{G}(2,1')
\nonumber \\
&&\hspace{8mm}
\pm 2 {\rm Tr}\check{G}(1,2')
\check{\tau}_{3}\check{G}(2',1')
\check{\tau}_{3}\check{G}(1',2)\check{G}(2,1)\bigr]
\, .
\label{Phi^(2)}
\end{eqnarray}
The self-energy $\check{\Sigma}^{(2)}$ is obtained
by eq.\ (\ref{Sigma-Phi}).
Expressing the result as a single matrix,
we observe that the elements 
of $\check{\Sigma}^{(2)}$ satisfy exactly the same relations as
$G_{ij}$ in eq.\ (\ref{G-prop}), in accordance with
eq.\ (\ref{checkSigma-prop}).
The off-diagonal elements are given by ($i\!\neq\! j$)
\begin{eqnarray}
&&\hspace{-15mm}
\Sigma_{ij}^{(2)}(1,1')=
\mp(\hbar)^{2} \!\int\! {\rm d}2 \!\int\! {\rm d}2'\,
\bar{V}(1\!-\!2)\bar{V}(1'\!-\!2')
\nonumber \\
&&\hspace{5mm}
\times\bigl[G_{ij}(1,1')G_{ji}(2',2)G_{ij}(2,2')
\nonumber \\
&&\hspace{7mm}
\pm G_{ij}(1,2')G_{ji}(2',2)G_{ij}(2,1')\bigr]\, .
\label{Sigma^(2)}
\end{eqnarray}
The symmetry of $\check{\Sigma}^{(2)}$ just mentioned is clearly a general property
of the higher-order contributions to $\check{\Sigma}$, 
as may be checked order by order.
Combining it with eq.\ (\ref{Sigma^(1)b}),
we now realize that $\Sigma_{11}$ and $\Sigma_{22}$,
which satisfy eq.\ (\ref{checkSigma-prop}),
can be written more specifically as 
\begin{subequations}
\label{Sigma-prop}
\begin{eqnarray}
&&\hspace{-10mm}
\Sigma_{11}(1,1')=\Sigma^{{\rm HF}}(1,1')
+\theta(t_{1}'\!-\!t_{1})\Sigma_{12}(1,1')
\nonumber \\
&&\hspace{8mm}+
\theta(t_{1}\!-\!t_{1}')\Sigma_{21}(1,1') \, ,
\label{Sigma-prop2}
\end{eqnarray}
\begin{eqnarray}
&&\hspace{-10mm}
\Sigma_{22}(1,1')=-\Sigma^{{\rm HF}}(1,1')+\theta(t_{1}\!-\!t_{1}')\Sigma_{12}(1,1')
\nonumber \\
&&\hspace{8mm}
+\theta(t_{1}'\!-\!t_{1})\Sigma_{21}(1,1') \, ,
\label{Sigma-prop3}
\end{eqnarray}
\end{subequations}
with $\Sigma^{{\rm HF}}$ given by eq.\ (\ref{Sigma-HF}).

\begin{figure}[t]
\begin{center}
  \includegraphics[width=0.8\linewidth]{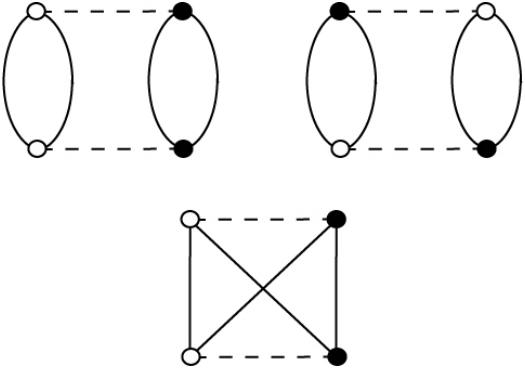}
\end{center}
  \caption{Topologically distinct second-order diagrams for $\Phi$.}
  \label{fig:4}
\end{figure}

Besides the one-particle Green's function, 
the $\Phi$-derivative approximation also provides us with a definite
and consistent evaluation scheme 
for higher-order correlations, as already shown by Baym and Kadanoff\cite{KB61}
and Baym\cite{Baym62} on the equilibrium imaginary-time contour.
However, this topic on the Keldysh contour seems not to have been paid 
due attention in the literature.
Especially important among them is the two-particle correlation function:
\begin{eqnarray}
&&\hspace{-5mm}
{\cal K}_{ij,kl}(12,34)\equiv \!\left(\!-\frac{i}{\hbar}\right)^{\!\!2}\!
\langle T_{C}
\psi_{{\cal H}}(1^{i})\psi_{{\cal H}}(3^{k})\psi_{{\cal H}}^{\dagger}(4^{l})
\psi_{{\cal H}}^{\dagger}(2^{j})\rangle 
\nonumber \\
&&\hspace{20mm} -G_{ij}(1,2)G_{kl}(3,4) \, .
\label{K}
\end{eqnarray}
Note that the arrangement of the space-time arguments in ${\cal K}$ is different
from that of Baym and Kadanoff;\cite{KB62}
the present one may be more convenient when regarding ${\cal K}$ as a matrix.

To consider higher-order correlations in a unified way,
we introduce an additional perturbation on the Keldysh contour
caused by the non-local one-body potential $W(1^{C},2^{C})$.
Adopting the interaction representation 
in terms of ${\cal H}$ in eq.\ (\ref{calH}),
the time evolution due to $W$ is described by the operator:
\begin{equation}
S'\equiv T_{C}\exp\!\left[-\frac{i}{\hbar}\int{\rm d}1\int{\rm d}1'\,
\vec{\psi}_{\cal H}^{\dagger}(1')\check{\tau}_{3}\check{W}(1',1)\check{\tau}_{3}
\vec{\psi}_{\cal H}(1)\right] ,
\label{S'}
\end{equation}
where $\check{\tau}_{3}$'s 
originate from the transformation (\ref{int_C}).
The Green's function is now given by
\begin{equation}
\check{G}(1,1';\check{W})=-\frac{i}{\hbar}\langle T_{C}S'
\vec{\psi}_{\cal H}(1)\vec{\psi}_{\cal H}^{\dagger}(1')\rangle/
\langle T_{C}S'\rangle
\, .
\end{equation}
The correlation function (\ref{K}) is then obtained by
\begin{equation}
{\cal K}_{ii',jj'}(11',22')=\!\left.\pm (-1)^{j+j'}
\frac{\delta G_{ii'}(1,1';\check{W})}{
\delta W_{j'j}(2',2)}\right|_{\check{W}=\check{0}} \, ,
\label{K2}
\end{equation}
where $\pm$ is due to the commutation relation 
between $\psi^{\dagger}$ and $\psi$, and
$(-1)^{j+j'}$ cancels the contribution of 
$\check{\tau}_{3}$'s in eq.\ (\ref{S'}).
To calculate ${\cal K}$ self-consistently 
in the $\Phi$-derivative approximation, 
we start from the Dyson equation (\ref{Dyson-check}).
It reads symbolically as $\check{G}^{-1}\check{G}\!=\!\check{1}$,
where $\check{G}^{-1}$ is given in the present case by
$\check{G}^{-1}\!=\!\check{\tau}_{3}(i\hbar\partial_{t}\!-\!K
\!-\!U)\!-\!\check{\tau}_{3}(\check{W}\!+\!\check{\Sigma})\check{\tau}_{3}$
with $K$ denoting the kinetic-energy operator.
Therefore, the first-order change $\delta\check{G}$ obeys
$\delta\check{G}\!=\!\check{G}(-\delta\check{G}^{-1})\check{G}$.
Let us substitute $-\delta\check{G}^{-1}\!=\!
\check{\tau}_{3}[\delta\check{W}+(\delta\check{\Sigma}/\delta\check{G})
(\delta \check{G}/\delta \check{W})\delta\check{W}]\check{\tau}_{3}$
into the first order equation and divide it by $\delta\check{W}$.
We thereby obtain an integral equation for eq.\ (\ref{K2}) as
\begin{eqnarray}
&&\hspace{-5mm}
{\cal K}_{ii',jj'}(11',22')=\pm G_{ij'}(1,2')G_{ji'}(2,1')
\nonumber \\
&&\hspace{-5mm}\pm i\hbar \sum_{kk'll'}
\int{\rm d}3\int{\rm d}4\int{\rm d}3'\int{\rm d}4' \, G_{ik'}(1,3')G_{ki'}(3,1')
\nonumber \\
&&\hspace{0mm}
\times \Gamma_{kk',ll'}(33',44') {\cal K}_{ll',jj'}(44',22') \, ,
\label{BS}
\end{eqnarray}
where $\Gamma$ is the irreducible vertex defined by
\begin{eqnarray}
&&\hspace{-12mm}
\Gamma_{ii',jj'}(11',22')\equiv 
\mp\frac{i}{\hbar}(-1)^{i+i'} \frac{\delta \Sigma_{i'i}(1',1)}{
\delta G_{jj'}(2,2')}
\nonumber \\
&&\hspace{12.2mm}
=-\frac{i}{\hbar}
\frac{\delta^{2}\Phi}{\delta G_{ii'}(1,1')\delta G_{jj'}(2,2')} \, .
\label{Xi}
\end{eqnarray}
We have used eq.\ (\ref{Sigma-Phi}) to derive the second expression
of eq.\ (\ref{Xi}).
Thus, once $\Phi$ is given explicitly as a functional of $\check{G}$,
the two-particle correlation (\ref{K}) can also be calculated 
by eqs.\ (\ref{BS}) and (\ref{Xi}).

The integral equation (\ref{BS}) may be solved iteratively
to obtain a formal solution:
\begin{equation}
\check{\check{\mbox{$\cal K$}}}=\pm \bigl(\,\check{\check{1}}\mp i\hbar\,
\check{G}\check{G}\,\check{\check{\Gamma}}\,\bigr)^{-1}\check{G}\check{G} \, ,
\label{K-2}
\end{equation}
where  
$\check{\check{1}}$ and $\check{G}\check{G}$ are matrices defined by
\begin{subequations}
\label{checkcheck}
\begin{equation}
(\check{\check{1}})_{ii',jj'}(11',22')=\delta_{ij}\delta_{i'j'}\delta(1,2)
\delta(1',2')\, ,
\end{equation}
\begin{equation}
(\check{G}\check{G})_{ii',jj'}(11',22')=G_{ij'}(1,2')G_{ji'}(2,1') \, ,
\label{GG}
\end{equation}
\end{subequations}
respectively.

Equation (\ref{Xi}) clearly has the symmetry:
\begin{subequations}
\label{Lambda-symm}
\begin{equation}
\Gamma_{ii',jj'}(11',22')=\Gamma_{jj',ii'}(22',11') \, .
\end{equation}
It also follows from eq.\ (\ref{checkG-prop}) that $\Gamma_{ii',jj'}$ satisfies
\begin{eqnarray}
&&\hspace{-8mm}
\Gamma_{ii',jj'}(11',22')
=-
\sum_{kk'll'}(\check{\tau}_{1})_{ik}(\check{\tau}_{1})_{k'i'}
(\check{\tau}_{1})_{jl}(\check{\tau}_{1})_{l'j'}
\nonumber \\
&&\hspace{20mm}
\times\,
[\Gamma_{k'k,l'l}(1'1,2'2)]^{*}\, .
\end{eqnarray}
\end{subequations}
The expressions of $\Gamma^{(n)}_{ii',jj'}(11',22')$
for $n\!=\! 1,2$ are given in Appendix\ref{App:Gamma} to see their structures
explicitly.

\subsection{Keldysh transformation}

As seen from eq.\ (\ref{G-prop}), the four elements of 
$\check{G}$ are not independent.
This redundancy in $\check{G}$ is removed by the 
following modified Keldysh transformation:\cite{LO75,RS86}
\begin{equation}
\check{G}^{\rm K}\equiv 
\check{L}\check{\tau}_{3}
\check{G}\check{L}^{\dagger} 
= 
\left[
\begin{array}{cc}
G^{\rm R} & G^{\rm K} \\
0 & G^{\rm A}
\end{array}
\right] ,
\label{check-G-K}
\end{equation}
where $\check{L}$ is defined by
$\check{L}\!\equiv\!\frac{1}{\sqrt{2}}(\check{1}-i\check{\tau}_{2})$
with $\check{\tau}_{2}$ denoting the second Pauli matrix.
Thus, the $21$ element of $\check{G}^{\rm K}$ vanishes, and 
the others also satisfy
\begin{subequations}
\label{G-RAK-symm}
\begin{equation}
[ G^{{\rm R}}(1,1')]^{*}
=G^{\rm A}(1',1)\, , \hspace{5mm}
[ G^{\rm K}(1,1')]^{*}
=-G^{\rm K}(1',1)\, .
\label{G-RAK-symm1}
\end{equation}
We hence realize that $G^{{\rm R}}$ and $G^{{\rm K}}$ form
an alternative set of two independent elements
in the Keldysh space.
Indeed, they are connected with the Kadanoff-Baym functions
$G_{12}$ and $G_{21}$ as
\begin{equation}
G^{{\rm R}}(1,1')=\theta(t_{1}\!-\!t_{1}')[G_{21}(1,1')\!-\!
G_{12}(1,1')] \, ,
\label{G-RAK-symm2}
\end{equation}
\begin{equation}
G^{{\rm K}}(1,1')=G_{12}(1,1')\!+\!
G_{21}(1,1') \, .
\label{G-RAK-symm3}
\end{equation}
\end{subequations}
It also follows from eqs.\ (\ref{checkSigma-prop}) and
(\ref{Sigma-prop}) that
$\check{\Sigma}^{{\rm K}}\!\equiv\! \check{L}\check{\tau}_{3}
\check{\Sigma}\check{L}^{\dagger}$ can be written as
\begin{equation}
\check{\Sigma}^{\rm K}\equiv 
\check{L}\check{\tau}_{3}
\check{\Sigma}\check{L}^{\dagger} 
= 
\left[
\begin{array}{cc}
\Sigma^{\rm R} & \Sigma^{\rm K} \\
0 & \Sigma^{\rm A}
\end{array}
\right] .
\label{check-Sigma-K}
\end{equation}
Its elements satisfy
\begin{subequations}
\label{Sigma-RAK-symm}
\begin{equation}
[ \Sigma^{{\rm R}}(1,1')]^{*}
=\Sigma^{\rm A}(1',1)\, , \hspace{5mm}
[ \Sigma^{\rm K}(1,1')]^{*}
=-\Sigma^{\rm K}(1',1)\, .
\label{Sigma-RAK-symm1}
\end{equation}
The quantities $\Sigma^{{\rm R}}$ and $\Sigma^{{\rm K}}$
are given in terms of $\Sigma^{{\rm HF}}$, $\Sigma_{12}$ and $\Sigma_{21}$ as
\begin{equation}
\Sigma^{{\rm R}}(1,1')\!=\!\Sigma^{{\rm HF}}(1,1')\!+\!
\theta(t_{1}\!-\!t_{1}')[\Sigma_{21}(1,1')\!-\!
\Sigma_{12}(1,1')] \, ,
\label{Sigma-RAK-symm2}
\end{equation}
\begin{equation}
\Sigma^{{\rm K}}(1,1')=\Sigma_{12}(1,1')\!+\!
\Sigma_{21}(1,1') \, .
\label{Sigma-RAK-symm3}
\end{equation}
\end{subequations}
It is worth pointing out that the Keldysh transformation
is not useful in the perturbation expansions of $\check{G}$ and
$\Phi$, since it obscures the basic symmetry of
them. The transformation should be carried out only after finishing the
expansion in terms of $\check{G}$.

Applying the Keldysh transformation to
eq.\ (\ref{Dyson-check}), we obtain the Dyson equation for 
$\check{G}^{{\rm K}}$ as
\begin{eqnarray}
&&\hspace{-10mm}
\left[i\hbar \frac{\partial }{\partial t_{1}}\!+\! 
\frac{\hbar^{2}\nabla_{1}^{2}}{2m}\!-\!U(1)
\right]\!
\check{G}^{\rm K}(1,1')
\nonumber \\
&&\hspace{-10mm}
-\int \check{\Sigma}^{\rm K}(1,2)
\check{G}^{\rm K}(2,1')\,{\rm d}2 
=\delta(1,1')\check{1} \, .
\label{Dyson-K-check}
\end{eqnarray}
Thus, the equation for the retarded 
function $G^{{\rm R}}$ is completely decoupled from that
of the Keldysh component $G^{{\rm K}}$.

Finally, it follows from eqs.\ (\ref{Xi}),
(\ref{check-G-K}) and (\ref{check-Sigma-K}) that
the variation $\delta\check{\Sigma}^{{\rm K}}$
is connected with $\delta\check{G}^{{\rm K}}$ as
\begin{eqnarray}
&&\hspace{-10mm}
\delta \Sigma_{i'i}^{{\rm K}}(1',1)
=\pm i\hbar \sum_{jj'}\int{\rm d}2\int{\rm d}2'\,
\Gamma_{ii',jj'}^{{\rm K}}(11',22')
\nonumber \\
&&\hspace{11mm}\times\delta G_{jj'}^{{\rm K}}(2,2') \, ,
\label{deltaSigma^K-deltaG^K}
\end{eqnarray}
with $\Gamma_{ii',jj'}^{{\rm K}}$ defined by
\begin{eqnarray}
&&\hspace{-8mm}
\Gamma_{ii',jj'}^{{\rm K}}(11',22')
\equiv 
\sum_{kk'll'}(\check{L}\check{\tau}_{3})_{ik}(\check{L}^{\dagger})_{k'i'}
(\check{L}\check{\tau}_{3})_{jl}(\check{L}^{\dagger})_{l'j'}
\nonumber \\
&&\hspace{20mm}
\times
\Gamma_{kk',ll'}(11',22')\, .
\label{LambdaK-Lambda}
\end{eqnarray}
We realize from eq.\ (\ref{Lambda-symm}) that
$\Gamma_{ii',jj'}^{{\rm K}}$ satisfies
\begin{subequations}
\label{Lambda^K-symm}
\begin{equation}
\Gamma_{ii',jj'}^{{\rm K}}(11',22')=\Gamma_{jj',ii'}^{{\rm K}}(22',11') \, ,
\end{equation}
\begin{eqnarray}
&&\hspace{-8mm}
\Gamma_{ii',jj'}^{{\rm K}}(11',22')
=-
\sum_{kk'll'}(i\check{\tau}_{2})_{ik}(i\check{\tau}_{2})_{k'i'}
(i\check{\tau}_{2})_{jl}(\check{i\tau}_{2})_{l'j'}
\nonumber \\
&&\hspace{20mm}
\times [\Gamma_{k'k,l'l}^{{\rm K}}(1'1,2'2)]^{*}
\, .
\end{eqnarray}
\end{subequations}
The quantity $\Gamma_{ii',jj'}^{{\rm K}}$ has the advantage 
that we only need to consider its $9$ elements
instead of $16$ in $\Gamma_{ii',jj'}$ due to the 
vanishing $21$ elements in $\check{\Sigma}^{{\rm K}}$ and $\check{G}^{{\rm K}}$.

\section{Gradient expansion}

The theoretical framework of \S\ref{sec:perturb} enables us a formally
exact microscopic treatment of nonequilibrium dynamical systems.
However, the coupled equations (\ref{Sigma-Phi}) and
(\ref{Dyson-K-check}) are still too difficult to solve practically.
It is desirable to reduce their complexity down to a tractable level
without loosing the physical essentials.
To this end, we here adopt the Wigner representation and subsequently
carry out the gradient expansion to eqs.\ (\ref{Sigma-Phi}) and
(\ref{Dyson-K-check}).
To be specific, the Wigner representation of $\check{G}(1,2)$
is defined through
\begin{equation}
\check{G}(1,2)=\int\frac{{\rm d}^{3}p\,{\rm d}\varepsilon}{(2\pi\hbar)^{4}}
\check{G}({\bm p}\varepsilon,{\bm r}_{12}t_{12})
{\rm e}^{i({\bm p}\cdot\bar{\bm r}_{12}-\varepsilon \bar{t}_{12})/\hbar} ,
\label{checkG-Wig}
\end{equation}
with ${\bm r}_{12}\!\equiv\!({\bm r}_{1}\!+\!{\bm r}_{2})/2$,
$t_{12}\!\equiv\!(t_{1}\!+\!t_{2})/2$, 
$\bar{\bm r}_{12}\!\equiv\!{\bm r}_{1}\!-\!{\bm r}_{2}$, and
$\bar{t}_{12}\!\equiv\!t_{1}\!-\!t_{2}$.
The gradient expansion denotes an expansion 
with respect to ${\bm r}t$
of $\check{G}({\bm p}\varepsilon,{\bm r}t)$.
It forms a well-established basis for the microscopic derivation
of various transport equations such as the quantum\cite{KB62} 
and classical Boltzmann equation.
The basic assumption is that the scales of the space-time inhomogeneity
are much longer than the microscopic scales to achieve the local 
equilibrium such as the mean-free path and the collision time.
This condition is well satisfied 
in most of nonequilibrium steady states without time evolution
such as Rayleigh-B\'enard convection.\cite{CH93,Chandrasekhar61,Busse78,Croquette89,
Koschmieder93,deBruyn96,BPA00}

\subsection{Spectral and distribution functions}

As seen in eq.\ (\ref{G-prop}), 
there are essentially two independent components in
$\check{G}$, i.e., 
$G_{12}$ and $G_{21}$.
We here introduce an alternative pair of 
independent components, i.e., the spectral function $A$ and 
the distribution function $\phi$,\cite{KB62,Ivanov00}
which turn out to be more convenient.
The spectral function $A$ is defined by
\begin{eqnarray}
&&\hspace{-10mm}
A(1,2)\equiv i\left[G_{21}(1,2)-G_{12}(1,2)\right]
\nonumber \\
&&\hspace{1.5mm}=
\frac{1}{\hbar}
\langle \psi_{\cal H}(1)\psi_{\cal H}^{\dagger}(2)
\!\mp\! \psi_{\cal H}^{\dagger}(2)\psi_{\cal H}(1)\rangle \, .
\label{A-def}
\end{eqnarray}
Let us expand $A$ as eq.\ (\ref{checkG-Wig}).
It then follows from the equal-time commutation relation of the field operators
that $A({\bm p}\varepsilon,{\bm r}t)$ satisfies the sum rule:
\begin{equation}
\int_{-\infty}^{\infty}\! 
A({\bm p}\varepsilon,{\bm r}t)\, {\rm d}\varepsilon
= 2\pi \, .
\label{A-sum}
\end{equation}
We also conclude from $A^{*}(1,2)\!=\!A(2,1)$ that
$A({\bm p}\varepsilon,{\bm r}t)$ is real.
We next introduce the distribution function $\phi$
directly in the Wigner representation as
\begin{eqnarray}
&&\hspace{-8mm}
\phi({\bm p}\varepsilon,{\bm r}_{12}t_{12})
\nonumber\\ 
&&\hspace{-8mm}
\equiv\frac{\displaystyle \int_{-\infty}^{\infty}\! {\rm d}\bar{t}_{12} \!
\int \!{\rm d}^{3}\bar{r}_{12} \,
\frac{1}{\hbar}\langle \psi^{\dagger}_{\cal H}(2)\psi_{\cal H}(1)\rangle
{\rm e}^{-i({\bm p}\cdot\bar{\bm r}_{12}-\varepsilon \bar{t}_{12})/\hbar} 
}
{A({\bm p}\varepsilon,{\bm r}_{12}t_{12})} 
\, .
\nonumber\\ 
\label{f-def}
\end{eqnarray}
We find from $A^{*}({\bm p}\varepsilon,{\bm r}t)\!=\!
A({\bm p}\varepsilon,{\bm r}t)$ and 
$\langle \psi_{\cal H}^{\dagger}(2)\psi_{\cal H}(1)\rangle^{*}$ $=
\langle \psi_{\cal H}^{\dagger}(1)\psi_{\cal H}(2)\rangle$ that
$\phi({\bm p}\varepsilon,{\bm r}t)$ is also real.
Thus, both $A$ and $\phi$ are real in the Wigner representation.
They form an alternative pair of
independent quantities in $\check{G}$.

In the equilibrium theory, $\phi$ is just the Bose/Fermi distribution function
so that we only need to calculate $A$.
For nonequilibrium systems, in contrast, 
we have to determine $A$ and $\phi$ simultaneously.
The derivation of the quantum transport 
equation\cite{KB62,Keldysh64,Rainer83,RS86,BM90,LF94,HJ98,Ivanov99,Ivanov00,Kita01}
amounts to integrating out the spectral function $A$,
which contains detailed information on the density of states,
to obtain a single equation for $\phi$.
However, we shall proceed without the approximation here.

\subsection{Wigner representation of $\check{G}$ and $\check{\Sigma}$}

The Wigner representation of $\check{G}$ can be written
in terms of $A$ and $\phi$.
Indeed, the off-diagonal elements are transformed into
\begin{subequations}
\label{G-Wig}
\begin{equation}
G_{12}({\bm p}\varepsilon,{\bm r}t)
=\mp i
A({\bm p}\varepsilon,{\bm r}t)\phi({\bm p}\varepsilon,{\bm r}t) \, ,
\label{G_12-Wig}
\end{equation}
\begin{equation}
G_{21}({\bm p}\varepsilon,{\bm r}t)
=-iA({\bm p}\varepsilon,{\bm r}t)[1\pm \phi({\bm p}\varepsilon,{\bm r}t)] \, .
\label{G_21-Wig}
\end{equation}
\end{subequations}
Using eqs.\ (\ref{G-RAK-symm})
and (\ref{G-Wig}), we also obtain the Wigner representations
of $G^{{\rm R}}$, $G^{{\rm A}}$ and $G^{{\rm K}}$ as
\begin{subequations}
\label{G^RK-Wig}
\begin{equation}
G^{{\rm R}}({\bm p}\varepsilon,{\bm r}t)=[G^{{\rm A}}({\bm p}\varepsilon,{\bm r}t)]^{*}
=
\int_{-\infty}^{\infty}\frac{{\rm d}\varepsilon'}{2\pi}\,
\frac{A({\bm p}\varepsilon',{\bm r}t)}{\varepsilon_{+}-\varepsilon'} \, ,
\label{G^R-Wig}
\end{equation}
\begin{equation}
G^{{\rm K}}({\bm p}\varepsilon,{\bm r}t)
=-iA({\bm p}\varepsilon,{\bm r}t)[1\pm 2\phi({\bm p}\varepsilon,{\bm r}t)] \, ,
\label{G^K-Wig}
\end{equation}
\end{subequations}
with $\varepsilon_{+}\!\equiv\!\varepsilon\!+\!i0_{+}$.
Since it may not cause any confusion,
we use the same symbols $G_{ij}$ in both the coordinate and the Wigner representations.
All the symbols
given below without arguments
belong to the Wigner representation.

Let us move on to the self-energies of eqs.\ (\ref{Sigma-Phi}) 
and (\ref{check-Sigma-K}).
Following Ivanov {\em et al}.,\cite{Ivanov00} 
we express the Wigner representations of their
independent components in a form similar to
the above expressions as
\begin{subequations}
\label{Sigma-Wig}
\begin{equation}
\Sigma_{12}({\bm p}\varepsilon,{\bm r}t)
=\mp i
A_{\Sigma}({\bm p}\varepsilon,{\bm r}t)\phi_{\Sigma}({\bm p}\varepsilon,{\bm r}t) \, ,
\label{Sigma_12-Wig}
\end{equation}
\begin{equation}
\Sigma_{21}({\bm p}\varepsilon,{\bm r}t)
=-iA_{\Sigma}({\bm p}\varepsilon,{\bm r}t)[1\pm \phi_{\Sigma}({\bm p}\varepsilon,{\bm r}t)] \, ,
\label{Sigma_21-Wig}
\end{equation}
\end{subequations}
and
\begin{subequations}
\label{Sigma^RK-Wig}
\begin{eqnarray}
&&\hspace{-10mm}
\Sigma^{{\rm R}}({\bm p}\varepsilon,{\bm r}t)
=[\Sigma^{{\rm A}}({\bm p}\varepsilon,{\bm r}t)]^{*}
\nonumber \\
&&\hspace{-10mm}
=\Sigma^{{\rm HF}}({\bm p},{\bm r}t)
+\int_{-\infty}^{\infty}\frac{{\rm d}\varepsilon'}{2\pi}\,
\frac{A_{\Sigma}({\bm p}\varepsilon',{\bm r}t)}{\varepsilon_{+}-\varepsilon'} \, ,
\label{Sigma^R-Wig}
\end{eqnarray}
\begin{equation}
\Sigma^{{\rm K}}({\bm p}\varepsilon,{\bm r}t)
=-iA_{\Sigma}({\bm p}\varepsilon,{\bm r}t)[1\pm 2\phi_{\Sigma}({\bm p}\varepsilon,{\bm r}t)] \, .
\label{Sigma^K-Wig}
\end{equation}
\end{subequations}
It follows from eq.\ (\ref{checkSigma-prop}) that $A_{\Sigma}$ and $\phi_{\Sigma}$
are also real.
They are functionals of $A$ and $\phi$ 
in the $\Phi$-derivative approximation.

The operator in the square bracket of eq.\ (\ref{Dyson-K-check})
can also be transformed into a matrix of the space-time coordinates 
by multiplying it by $\delta(1,3)$ from the right.
Its Wigner representation is given by
\begin{equation}
G_{0}^{-1}({\bm p}\varepsilon,{\bm r}t)
\equiv\varepsilon-\frac{p^{2}}{2m}-U({\bm r}t) \, .
\label{M_0}
\end{equation}

\subsection{Gradient expansion}

We now consider the matrix product of the space-time coordinate.
It is transformed into the Wigner representation as\cite{RS86}
\begin{eqnarray}
&&\hspace{-10mm}
\int_{-\infty}^{\infty}\! {\rm d}\bar{t}_{12} \!
\int \!{\rm d}^{3}\bar{r}_{12} \,{\rm e}^{-i({\bm p}\cdot
\bar{\bm r}_{12}-\varepsilon\bar{t}_{12})/\hbar}
\int {\rm d}3 \, C(1,3)D(3,2) 
\nonumber \\
&&\hspace{-10mm}
=
C({\bm p}\varepsilon,{\bm r}_{12}t_{12})
\otimes
D({\bm p}\varepsilon,{\bm r}_{12}t_{12}) \, ,
\label{Identity-otimes}
\end{eqnarray}
where the operator $\otimes$ is defined by
\begin{subequations}
\label{otimes}
\begin{eqnarray}
&&\hspace{-10mm}
C({\bm p}\varepsilon,{\bm r}t)\!\otimes\!
D({\bm p}\varepsilon,{\bm r}t)
\nonumber \\
&&\hspace{-13mm}
\equiv
\exp\!\left[\frac{i\hbar}{2}
({\bm \partial}_{{\bm r}}\!\cdot\!{\bm \partial}_{{\bm p}'}
-\partial_{t}\partial_{\varepsilon'}
-{\bm \partial}_{{\bm p}}\!\cdot\!{\bm \partial}_{{\bm r}'}
+\partial_{\varepsilon}\partial_{t'})
\right]\!
\nonumber \\
&&\hspace{-10mm}
\times C({\bm p}\varepsilon,{\bm r}t)
D({\bm p}'\varepsilon',{\bm r}'t')\biggr|_{
{\bm p}'={\bm p},\varepsilon'=\varepsilon,{\bm r}'={\bm r},t'=t}\, ,
\label{otimes-t}
\end{eqnarray}
with ${\bm \partial}_{{\bm r}}\!\equiv\!\partial/\partial{\bm r}$
and $\partial_{t}\!\equiv\!\partial/\partial t$.
The identity (\ref{Identity-otimes}) can be proved as follows:
write 
$C$ and $D$ on the left-hand side
in the Wigner representation; expand ${\bm r}_{13}t_{13}$ and
${\bm r}_{32}t_{32}$ from ${\bm r}_{12}t_{12}$; 
remove $\bar{\bm r}_{13}\bar{t}_{13}$ and 
$\bar{\bm r}_{32}\bar{t}_{32}$ by using 
$\bar{\bm r}_{13}\,{\rm e}^{i{\bm p}'\cdot\bar{\bm r}_{13}/\hbar}
=-i\hbar\frac{\partial}{\partial {\bm p}'}
{\rm e}^{i{\bm p}'\cdot\bar{\bm r}_{13}/\hbar}$, etc.;
perform partial integrations over internal momentum-energy variables;
carry out the integrations over $3$ and 
$\bar{\bm r}_{12}\bar{t}_{12}$.
The first-order approximation to eq.\ (\ref{otimes-t}) yields
\begin{equation}
C\!\otimes\! D
\approx CD+\frac{i\hbar}{2}\{C,D\} \, ,
\label{otimes1}
\end{equation}
\end{subequations}
where the curly bracket denotes the generalized Poisson
bracket:
\begin{equation}
\{C,D\}\equiv\frac{\partial C}{\partial{\bm r}}\!\cdot\!
\frac{\partial D}{\partial{\bm p}}
-\frac{\partial C}{\partial t}\,
\frac{\partial D}{\partial\varepsilon}
-\frac{\partial C}{\partial{\bm p}}\!\cdot\!
\frac{\partial D}{\partial{\bm r}}
+\frac{\partial C}{\partial \varepsilon}\,
\frac{\partial D}{\partial t} \, .
\end{equation}
Finally, it should be pointed out that both the
Wigner transformation (\ref{checkG-Wig}) and the gradient expansion 
(\ref{Identity-otimes}) need
essential modifications in the presence 
of the electromagnetic field.
Here we need a special care on the gauge invariance of the
equations in order to appropriately obtain (i) the Hall terms\cite{HJ98,LF94,Kita01} 
and (ii) the pair potential of superconductivity as an effective
wave function of charge $2e$.\cite{Kita01}
To be specific, the Wigner transformation (\ref{checkG-Wig}) 
should be defined as eq.\ (7) of ref.\ \onlinecite{LF94} or
eq.\ (21) of ref.\ \onlinecite{Kita01}, 
and eq.\ (\ref{Identity-otimes}) above has to be replaced by 
eq.\ (36) of ref.\ \onlinecite{Kita01}.

\subsection{Dyson equation in the Wigner representation}

We now transform the Dyson equation into the Wigner representation
within the first-order gradient expansion.
Following Keldysh,\cite{Keldysh64} we start from eq.\ (\ref{Dyson-K-check}) 
rather than eq.\ (\ref{Dyson-check})
adopted by Kadanoff and Baym\cite{KB62} and
Ivanov {\em et al}.\cite{Ivanov00}
Indeed, eq.\ (\ref{Dyson-K-check}) has the advantages that
(i) its $21$ element vanishes
and (ii) the $22$ element is complex-conjugate of the $11$ element.
Hence we essentially need to consider only
the first row of eq.\ (\ref{Dyson-K-check}),
which completely determines the two independent components
of $\check{G}$, i.e.,  $A$ and $\phi$.
Thus, we can see the structure of the equations more clearly
in the present approach.

Using eq.\ (\ref{Identity-otimes}) and (\ref{otimes1}),
the $11$ element of eq.\ (\ref{Dyson-K-check})
is transformed into
\begin{subequations}
\label{Dyson-R-Wig}
\begin{equation}
(G_{0}^{-1}\!-\!\Sigma^{{\rm R}})G^{{\rm R}}
+\frac{i\hbar}{2}\{G_{0}^{-1}\!-\!\Sigma^{{\rm R}},G^{{\rm R}}\}
=1 \, ,
\label{Dyson-Rl-Wig}
\end{equation}
where $G_{0}^{-1}$ is defined by eq.\ (\ref{M_0}).
The replacement ${\rm R}\!\rightarrow\!{\rm A}$
in the superscript
yields the equation for the $22$ element.
Taking its complex conjugate and noting
eq.\ (\ref{G^R-Wig}), we have an alternative equation for
$G^{{\rm R}}$ as
\begin{equation}
(G_{0}^{-1}\!-\!\Sigma^{{\rm R}})G^{{\rm R}}
-\frac{i\hbar}{2}\{G_{0}^{-1}\!-\!\Sigma^{{\rm R}},G^{{\rm R}}\}
=1 \, .
\label{Dyson-Rr-Wig}
\end{equation}
\end{subequations}
Let us add eqs.\ (\ref{Dyson-Rl-Wig}) and (\ref{Dyson-Rr-Wig}).
We then obtain
\begin{equation}
G^{{\rm R}}=(G_{0}^{-1}-\Sigma^{{\rm R}})^{-1} \, .
\label{G^R}
\end{equation}
We realize from eq.\ (\ref{G^R-Wig}) that this is the equation to
determine the spectral function $A$ for a given $\Sigma^{{\rm R}}$.
Since the $11$ and $22$ elements of
eq.\ (\ref{Dyson-K-check}) were equivalent before the gradient expansion, 
it is natural to ask 
whether this property is still retained between eqs.\ (\ref{Dyson-Rl-Wig})
and (\ref{Dyson-Rr-Wig}). To answer the question, let us subtract eq.\
(\ref{Dyson-Rr-Wig}) from eq.\ (\ref{Dyson-Rl-Wig})
and substitute eq.\ (\ref{G^R}) in the resulting equation.
We then obtain
$
0=\{(G^{{\rm R}})^{-1},G^{{\rm R}}\}=
-(G^{{\rm R}})^{-2}\{G^{{\rm R}},G^{{\rm R}}\}$, which is just $0\!=\!0$.
We have thereby confirmed the equivalence between eqs.\ (\ref{Dyson-Rl-Wig})
and (\ref{Dyson-Rr-Wig}).
We now realize that $A$ can be determined locally by eq.\ (\ref{G^R})
without space-time derivatives
within the first-order gradient expansion.

It follows from the retarded nature of $G^{{\rm R}}(1,2)$ in eq.\ (\ref{G-RAK-symm2})
that all the singularities of $G^{{\rm R}}({\bm p}\varepsilon,{\bm r}t)$
in eq.\ (\ref{G^R}) lie on the lower half of the complex $\varepsilon$ plane.
This implies ${\rm Im}\Sigma^{{\rm R}}({\bm p}\varepsilon,{\bm r}t)
\!\leq\! 0$ and hence 
${\rm Im}G^{{\rm R}}({\bm p}\varepsilon,{\bm r}t)\!\leq\! 0$. 
Using eq.\ (\ref{G^R-Wig}), the latter condition can be written 
alternatively as
\begin{equation}
A({\bm p}\varepsilon,{\bm r}t)\geq 0 \, .
\end{equation}

Next, the gradient expansion to the $12$ element
of eq.\ (\ref{Dyson-K-check}) leads to
\begin{subequations}
\label{Dyson-K-Wig}
\begin{eqnarray}
&&\hspace{-10mm}
(G_{0}^{-1}\!-\!\Sigma^{{\rm R}})G^{{\rm K}}-\Sigma^{{\rm K}}G^{{\rm A}}
+\frac{i\hbar}{2}\{G_{0}^{-1}\!-\!\Sigma^{{\rm R}},G^{{\rm K}}\}
\nonumber \\
&&\hspace{-10mm}
-\frac{i\hbar}{2}\{\Sigma^{{\rm K}},G^{{\rm A}}\}
=0 \, .
\label{Dyson-Kl-Wig}
\end{eqnarray}
Taking its complex conjugate
and using eqs.\ (\ref{G^RK-Wig}) and (\ref{Sigma^RK-Wig}),
we have an alternative expression of eq.\ (\ref{Dyson-Kl-Wig}) as
\begin{eqnarray}
&&\hspace{-10mm}
(G_{0}^{-1}\!-\!\Sigma^{{\rm A}})G^{{\rm K}}-\Sigma^{{\rm K}}G^{{\rm R}}
-\frac{i\hbar}{2}\{G_{0}^{-1}\!-\!\Sigma^{{\rm A}},G^{{\rm K}}\}
\nonumber \\
&&\hspace{-10mm}
+\frac{i\hbar}{2}\{\Sigma^{{\rm K}},G^{{\rm R}}\}
=0 \, .
\label{Dyson-Kr-Wig}
\end{eqnarray}
\end{subequations}
Let us subtract eq.\ (\ref{Dyson-Kr-Wig}) from eq.\ (\ref{Dyson-Kl-Wig}),
substitute eqs.\ (\ref{G^RK-Wig}) and (\ref{Sigma^RK-Wig})
into the resulting equation, and use 
$\{G_{0}^{-1}\!-\!\Sigma^{{\rm R}},G^{{\rm R}}\}\!=\!0$.
We thereby obtain
$$
\{G_{0}^{-1}\!-\!{\rm Re}\Sigma^{{\rm R}},A \phi\}-\{A_{\Sigma}\phi_{\Sigma},{\rm Re}G^{{\rm R}}\}
=\frac{AA_{\Sigma}(\phi_{\Sigma}-\phi)}{\hbar}\, .
$$
The left-hand side of the equation consists of terms with space-time
derivatives, whereas the right-hand side denotes the collision integral which
vanishes in equilibrium.
Hence it is appropriate in the first-order gradient expansion
to replace $\phi_{\Sigma}$ on the left-hand side by $\phi$.
The approximation was originally suggested by Botermans and Malfliet\cite{BM90}
and adopted explicitly by Ivanov {\em et al}.\cite{Ivanov00}
The idea also has a close relationship with the Enskog series
for solving the Boltzmann equation,\cite{CC90,HCB54}
i.e., the expansion from the local equilibrium.
The procedure leads to
\begin{equation}
\{G_{0}^{-1}\!-\!{\rm Re}\Sigma^{{\rm R}},A \phi\}-\{A_{\Sigma} \phi,{\rm Re}G^{{\rm R}}\}
={\cal C}\, ,
\label{transport}
\end{equation}
where ${\cal C}$ denotes the collision integral defined by
\begin{equation}
{\cal C}\equiv \frac{AA_{\Sigma}(\phi_{\Sigma}-\phi)}{\hbar}=
\mp\frac{G_{21}\Sigma_{12}-G_{12}\Sigma_{21}}{\hbar} \, .
\label{collision}
\end{equation}
Equation (\ref{transport}) determines $\phi$ for given $A$ and $\check{\Sigma}$.

Following Ivanov {\em et al}.,\cite{Ivanov00} we now ask the question of whether eqs.\
(\ref{Dyson-Kl-Wig}) and (\ref{Dyson-Kr-Wig}) still hold the equivalence
which was present before the gradient expansion between the $12$ element of eq.\
(\ref{Dyson-K-check}) and its complex conjugate.
Let us add eqs.\
(\ref{Dyson-Kl-Wig}) and (\ref{Dyson-Kr-Wig}).
After the same procedures as above for the subtraction, we obtain
\begin{eqnarray*}
&&\hspace{-10mm}
\frac{\{A_{\Sigma} \phi,A\}-\{A_{\Sigma},A \phi\}}{4}
\nonumber \\
&&\hspace{-10mm}
=\frac{A_{\Sigma} \phi_{\Sigma}{\rm Re}G^{{\rm R}}
-(G_{0}^{-1}\!-\!{\rm Re}\Sigma^{{\rm R}})A \phi}{\hbar} \, .
\end{eqnarray*}
This equation is identical with eq.\ (\ref{transport}).
Indeed, multiplying it by $A_{\Sigma}/(G_{0}^{-1}\!-\!{\rm Re}\Sigma^{{\rm R}})$
yields eq.\ (\ref{transport}) in disguise.
This may be seen more explicitly 
by (i) expressing ${\rm Re}G^{{\rm R}}$ and $A$ 
in the two apparently different equations with respect to 
${\rm Re}G^{{\rm R}-1}\!\equiv\!G_{0}^{-1}\!-\!{\rm Re}\Sigma^{{\rm R}}$ and 
$A_{\Sigma}$, and (ii) transforming the gradient terms
into derivatives of ${\rm Re}G^{{\rm R}-1}$, $A_{\Sigma}$ and $\phi$.\cite{Ivanov00}
Thus, with the approximation $\phi_{\Sigma}\!\rightarrow\! \phi$
in the gradient terms, eqs.\ (\ref{Dyson-Kl-Wig}) and (\ref{Dyson-Kr-Wig}) 
recover the original equivalence.

Equations (\ref{G^R}) and (\ref{transport}) form coupled equations
to completely determine the real quantities $A$ and $\phi$ for 
a given $\check{\Sigma}$.
Indeed, eq.\ (\ref{transport}) is real, and the real and imaginary parts of 
eq.\ (\ref{G^R}) are connected by the Kramers-Kronig relation.

In Appendix\ref{app:conserve-W}, 
we derive basic conservation laws
in the first-order gradient expansion of the $\Phi$-derivative
approximation.

\subsection{Self-energy in the Wigner representation}
\label{subsec:grad-self}

The original $\Phi$-derivative approximation denotes solving
eqs.\ (\ref{Sigma-Phi}) and (\ref{Dyson-K-check}) self-consistently.
Having performed the first-order gradient expansion 
to the Dyson equation (\ref{Dyson-K-check}), 
we also have to specify a consistent
approximation scheme to the other equation (\ref{Sigma-Phi}).
This issue seems not to have been given an explicit consideration before, 
however.

As noted below it,
eq.\ (\ref{G^R}) is correct up to first order in the gradient expansion.
This implies that the local approximation to $\Sigma^{{\rm R}}$ 
is sufficient for solving eq.\ (\ref{G^R}).
As for eq.\ (\ref{transport}), 
all the terms on the left-hand side include space-time derivatives,
and the collision term of the right-hand side 
is connected by equality with the left-hand side.
Thus, eq.\ (\ref{transport}) is a first-order equation
where every quantity should be evaluated locally
without derivatives.
With these considerations,
we now conclude that we have to 
apply the local approximation to eq.\ 
(\ref{Sigma-Phi}).
We have thereby reached the definite prescription
to evaluate $\check{\Sigma}$ in terms of $\check{G}$, so that
eqs.\ (\ref{G^R}) and (\ref{transport}) now 
form closed equations for $A$ and $\phi$.

Using eq.\ (\ref{V-Fourier}),
the Hartree-Fock self-energy (\ref{Sigma-HF}) is 
transformed into
the Wigner representation as
\begin{equation}
\Sigma^{{\rm HF}}({\bm p},{\bm r}t)=\pm i\hbar
\int\! \frac{{\rm d}^{3}p'{\rm d}\varepsilon'}{(2\pi \hbar)^{4}}
(V_{0}\pm V_{{\bm p}-{\bm p}'}) 
G_{12}({\bm p}'\varepsilon',{\bm r}t) \, .
\label{Sigma-HF-Wig}
\end{equation}
Also, the local approximation to eq.\ (\ref{Sigma^(2)}) 
yields ($i\!\neq\! j$)
\begin{eqnarray}
&&\hspace{-8mm}
\Sigma_{ij}^{(2)}({\bm p}\varepsilon,{\bm r}t)
=\mp (\hbar)^{2} \! 
\int\!\prod_{k=2}^{4}\frac{{\rm d}^{3}p_{k}{\rm d}\varepsilon_{k}}{(2\pi\hbar)^{4}}
\frac{1}{2}|V_{{\bm p}-{\bm p}_{3}}\!\pm\!V_{{\bm p}-{\bm p}_{4}}|^{2}
\nonumber \\
&&\hspace{0mm}
\times (2\pi\hbar)^{4}
\delta({\bm p}\!+\!{\bm p}_{2}\!-\!{\bm p}_{3}\!-\!{\bm p}_{4})
\delta(\varepsilon\!+\!\varepsilon_{2}\!-\!\varepsilon_{3}
\!-\!\varepsilon_{4})
\nonumber \\
&&\hspace{0mm}
\times G_{ji}({\bm p}_{2}\varepsilon_{2},{\bm r}t)
G_{ij}({\bm p}_{3}\varepsilon_{3},{\bm r}t)
G_{ij}({\bm p}_{4}\varepsilon_{4},{\bm r}t)\, .
\label{Sigma^(2)-Wig}
\end{eqnarray}
Writing $G_{12}\!=\!\mp iG^{<}$, $G_{21}\!=\!- iG^{>}$,
$\Sigma_{12}\!=\!\mp i\Sigma^{<}$ and $\Sigma_{21}\!=\!- i\Sigma^{>}$
in the Wigner representation, we find that eq.\ (\ref{Sigma^(2)-Wig}) 
is identical with eq.\ (4-16) of Kadanoff and Baym,\cite{KB62} 
as they should.

Contrary to the local approximation adopted here,
Ivanov {\em et al}.\ \cite{Ivanov00} emphasized the importance of considering
the first-order gradient corrections to the collision integral,
which they call memory corrections.
Their motivation towards this conclusion seems to be stemming from
the expression of equilibrium entropy obtained by Carneiro and Pethick;\cite{CP75}
see the paragraph at the end in \S 5.4 of
Ivanov {\em et al}.\ \cite{Ivanov00}
Indeed, they have shown that the Carneiro-Pethick expression cannot be 
reproduced from their dynamical equations without the memory
corrections.
On the other hand, we have already seen that the local approximation 
should be sufficient
for the collision integral.
Thus, one may wonder which statement is correct and 
where the discrepancy originates from.
We will show that: (i) the Carneiro-Pethick expression of equilibrium entropy is
not correct due to an inappropriate treatment of energy denominators
in their calculation of the thermodynamic potential;
and (ii) the local approximation without the memory corrections 
leads to an expression of dynamical entropy which is compatible
with the equilibrium expression.\cite{Kita99}
Thus, the memory corrections should not be incorporated within the first-order
gradient expansion.

\section{Entropy}

We are now ready to discuss entropy in nonequilibrium statistical mechanics.
We first derive an equation of motion for entropy density in \S\ref{subsec:dsdt}.
We then prove the $H$-theorem for a limiting case in \S\ref{subsec:H-theorem}.
Finally, we propose a principle of maximum entropy for nonequilibrium 
steady states in \S\ref{subsec:Smax}

\subsection{\label{subsec:dsdt}Equation of motion for entropy density}
Let us multiply eq.\ (\ref{transport}) by $k_{\rm B}\ln[(1\!\pm\! \phi)/\phi]$
and integrate it over ${\bm p}$ and $\varepsilon$.
We next write $k_{\rm B}\ln[(1\pm \phi)/\phi]{\rm d}\phi\!=\! {\rm d}\sigma$
in the resulting equation, where
$\sigma$ denotes entropy of the noninteracting system:\cite{LL80}
\begin{equation}
\sigma\equiv k_{\rm B}[-\phi\ln\phi\pm (1\pm \phi)\ln (1\pm \phi)] \, .
\label{sigma}
\end{equation}
Thus, the derivatives of $\phi$ on the left-hand side are transformed
into those of $\sigma$.
We then perform partial integrations over ${\bm p}$ and $\varepsilon$.
We thereby arrive at an equation of motion:
\begin{equation}
\frac{\partial s}{\partial t}+{\bm \nabla}\cdot{\bm j}_{s}=k_{\rm B}\hbar
\!\int \!\frac{{\rm d}^{3}p\,{\rm d}\varepsilon}{(2\pi\hbar)^{4}}\,
{\cal C}\ln\frac{1\pm \phi}{\phi} \, ,
\label{continuity-s}
\end{equation}
with $s\!=\!s({\bm r}t)$ and ${\bm j}_{s}\!=\!{\bm j}_{s}({\bm r}t)$ defined by
\begin{equation}
s\equiv \hbar \int \frac{{\rm d}^{3}p\,{\rm d}\varepsilon}{(2\pi\hbar)^{4}}\,
\sigma\!\left[A\frac{\partial (G_{0}^{-1}\!-\!{\rm Re}\Sigma^{{\rm R}})}{\partial \varepsilon}
+A_{\Sigma} \frac{\partial {\rm Re}G^{{\rm R}}}{\partial \varepsilon}\right] ,
\label{entropy-density}
\end{equation}
\begin{equation}
{\bm j}_{s}\equiv \hbar \int \frac{{\rm d}^{3}p\,{\rm d}\varepsilon}{(2\pi\hbar)^{4}}\,
\sigma\!\left[-A\frac{\partial (G_{0}^{-1}\!-\!{\rm Re}\Sigma^{{\rm R}})}{\partial {\bm p}}
-A_{\Sigma} \frac{\partial {\rm Re}G^{{\rm R}}}{\partial {\bm p}}\right] .
\label{entropy-current}
\end{equation}
Let us explain the quantities in these expressions once again for an easy reference.
The quantity ${\cal C}$ is the collision integral (\ref{collision}),
$\phi$ the distribution function defined by eq.\ (\ref{f-def}),
$A$ the Wigner transformation of eq.\ (\ref{A-def}) called the spectral function,
$G_{0}^{-1}$ defined by eq.\ (\ref{M_0}), and the retarded functions
$G^{{\rm R}}$ and $\Sigma^{{\rm R}}$ 
given by eqs.\ (\ref{G^R-Wig}) and (\ref{Sigma^R-Wig}), respectively.
The basic quantities $A$ and $\phi$ should be determined self-consistently by eqs.\
(\ref{G^R}) and (\ref{transport}) 
using the self-energy of the local approximation,
as discussed in detail in \S\ref{subsec:grad-self}.

The right-hand side of eq.\ (\ref{continuity-s}) denotes
net change of entropy due to collisions, 
whereas $s$ and ${\bm j}_{s}$ on the left-hand side
can be regarded as the entropy density 
and the entropy flux density, respectively.
Indeed, eq.\ (\ref{entropy-density}) agrees completely in form
with the equilibrium expression of entropy, i.e.,
eq.\ (5) of ref.\ \onlinecite{Kita99}.\cite{comment2}
This may be shown explicitly from
the latter
by: (i) noting $\partial \phi/\partial T=-{\partial\sigma}/{
\partial \varepsilon}$ for $\phi\!=\!({\rm e}^{\varepsilon/k_{\rm B}T}\! \pm\! 1)^{-1}$
in equilibrium;
and (ii) performing a partial integration with respect to
$\varepsilon$.
Thus, eq.\ (\ref{entropy-density}) is compatible with
equilibrium statistical mechanics and may be regarded as an
expression of the nonequilibrium entropy density.

Equation (\ref{entropy-density}) was obtained by
Ivanov {\em et al}.\ \cite{Ivanov00} as the equation
for ``Markovian entropy flow.'' 
However, they claim that there is additional contribution to entropy
called ``memory effects,'' 
which originates from the gradient terms in the collision
integral on the right-hand side of eq.\ (\ref{transport}).
Note that the gradient terms in the collision
integral have been discarded in the present formulation
with rationales given in the second paragraph of \S\ref{subsec:grad-self}.
By including the memory effects, Ivanov {\em et al}.\ \cite{Ivanov00} could obtain
an expression of entropy which is compatible with the equilibrium
entropy derived earlier by Carneiro and Pethick.\cite{CP75}

However, Carneiro and Pethick\cite{CP75} obtained their expression
with the zero-temperature time-ordered Goldstone technique
where there may be ambiguity as to how to deal with vanishing
energy denominators.
Indeed, Carneiro and Pethick added an infinitesimal imaginary quantity
to every energy denominator in their calculation of $\Phi$ in equilibrium.
They thereby found a contribution to $\Phi$
from singularities in the energy denominators, i.e., 
the ``on-energy-shell'' term;
see the arguments in \S IV.A of their paper.\cite{CP75}
It is this on-energy-shell contribution to $\Phi$ which brings
the difference between eq.\ (\ref{entropy-density})
and the Carneiro-Pethick expression.

Since the issue has a crucial importance to the whole theory,
we have reexamined in Appendix\ref{app:Phi}
whether such singular contribution to $\Phi$
is really present or not.
To this end, we adopt the finite-temperature
formalism of using the Matsubara Green's function.
A definite advantage of the present approach is that
no additional regularization procedure is necessary.
It is thereby shown that the on-energy-shell contribution is absent.
Thus, it is eq.\ (\ref{entropy-density}) which is compatible
with the equilibrium expression of entropy.
The conclusion also implies that we need not consider the 
``memory effects'' of Ivanov {\em et al}.\cite{Ivanov00}

\subsection{\label{subsec:H-theorem}Entropy production and the $H$-theorem}

As it has already been mentioned, 
the right-hand side of eq.\ (\ref{continuity-s}) expresses
net change of entropy due to collisions with ${\cal C}$ given by
eq.\ (\ref{collision}).
We hence put
\begin{equation}
\mp k_{\rm B}
\int \!\frac{{\rm d}^{3}p\,{\rm d}\varepsilon}{(2\pi\hbar)^{4}}\,
(G_{21}\Sigma_{12}-G_{12}\Sigma_{21})\ln\frac{1\pm \phi}{\phi} 
\equiv \frac{\partial s_{\rm col}}{\partial t} \, ,
\label{s_col}
\end{equation}
and study this term more closely.
It is shown shortly below that eq.\ (\ref{s_col}) is positive within the second-order
perturbation expansion.

Let us substitute eq.\ (\ref{Sigma^(2)-Wig}) into eq.\ (\ref{s_col})
and rewrite it in terms of 
the Kadanoff-Baym functions defined in the Wigner representation by\cite{KB62}
$G^{<}\!\equiv\!\pm i G_{12}\!=\!A\phi$ and $G^{>}\!\equiv\!
i G_{21}\!=\!A(1\!\pm \phi)$.
Equation (\ref{s_col}) is thereby transformed into
\begin{eqnarray}
&&\hspace{-5mm}
\frac{\partial s_{\rm col}^{(2)}}{\partial t}
= k_{\rm B}\hbar^{2} \int \prod_{j=1}^{4}
\frac{{\rm d}^{3}p_{j}\,{\rm d}\varepsilon_{j}}{(2\pi\hbar)^{4}}
\frac{1}{2}|V_{{\bm p}_{1}-{\bm p}_{3}}\!\pm\!V_{{\bm p}_{1}-{\bm p}_{4}}|^{2}
\nonumber \\
&&\hspace{1mm}
\times(2\pi\hbar)^{4}
\delta({\bm p}_{1}\!+\!{\bm p}_{2}\!-\!{\bm p}_{3}\!-\!{\bm p}_{4})
\delta(\varepsilon_{1}\!+\!\varepsilon_{2}\!-\!\varepsilon_{3}\!-\!\varepsilon_{4})
\nonumber \\
&&\hspace{1mm}
\times
\bigl[G^{<}_{1}G^{>}_{2}G^{<}_{3}G^{>}_{4}-G^{>}_{1}G^{<}_{2}G^{>}_{3}G^{<}_{4}\bigr]
\ln\frac{G^{<}_{1}G^{>}_{2}G^{<}_{3}G^{>}_{4}}{G^{>}_{1}G^{<}_{2}G^{>}_{3}G^{<}_{4}}
\, ,
\nonumber \\
&& \label{ds_col/dt}
\end{eqnarray}
with $G^{<}_{j}\!\equiv\!G^{<}({\bm p}_{j}\varepsilon_{j},{\bm r}t)$ and
$G^{>}_{j}\!\equiv\!G^{>}({\bm p}_{j}\varepsilon_{j},{\bm r}t)$.
Using the inequality $(x\!-\!y)\ln(x/y)\!\geq\! 0$ which holds
for any positive $x$ and $y$,\cite{Cercignani88,Cercignani98} we then conclude
\begin{equation}
{\partial s_{\rm col}^{(2)}}({\bm r}t)/{\partial t}\geq 0 \, .
\label{s_col^(2)}
\end{equation}
Thus, entropy increases by collision in the second-order
perturbation.
This is quite a strong statement in that the inequality holds even locally.

The $H$-theorem is relevant to the 
space integral of eq.\ (\ref{continuity-s}) 
over the whole system:\cite{Cercignani88,Cercignani98}
\begin{equation}
\frac{{\rm d}}{{\rm d} t} \int s({\bm r}t)\, {\rm d}^{3}r 
+\int {\bm j}_{s}({\bm r}t)\cdot {\rm d}{\bm S}
=  \frac{{\rm d}}{{\rm d} t}\int s_{\rm col}({\bm r}t)\, {\rm d}^{3}r \, ,
\label{s-flow-tot}
\end{equation}
where ${\rm d}{\bm S}$ denotes the infinitesimal surface element.
We review it here for a later extension:
Consider an isolated system where the second term on the left-hand side
vanishes. 
If the time evolution of entropy by collision is globally positive,
i.e., 
\begin{equation}
\frac{{\rm d}}{{\rm d} t} \int s_{\rm col}({\bm r}t)\, {\rm d}^{3}r\geq 0 \, ,
\label{s_col>0}
\end{equation}
we obtain the law of increase of entropy
for the relevant system as
\begin{equation}
\frac{{\rm d}}{{\rm d} t} \int s({\bm r}t)\, {\rm d}^{3}r\geq 0 \, .
\label{s>0}
\end{equation}
It hence follows that entropy takes its maximum value
in equilibrium of an isolated system.

There seems to have been yet no explicit proof for eq.\ (\ref{s_col>0}) 
beyond the second-order perturbation.
Indeed, the $H$-theorem has been discussed almost exclusively
in terms of the Boltzmann equation with the two-particle (i.e., second-order) 
collision integral.
A general expression corresponding to eq.\ (\ref{ds_col/dt}) has been 
provided for the contact interaction as eq.\ (5.11) of Ivanov {\em et al}.\cite{Ivanov00}
A sufficient condition for the general $H$-theorem to hold is
$R_{m,m'}\!\geq\! 0$ in their expression at every order of the perturbation expansion,
which seems nontrivial to prove at present 
and we defer it for a future study.
However, it should be noted that 
Nature clearly adopts the inequality (\ref{s_col>0}) beyond the second-order,
since no explicit violation of the second law of thermodynamics 
has been reported even in systems with strong correlations.
See also the numerical work by Orban and Bellemans 
on the time evolution of entropy of an isolated system.\cite{OB67}

\subsection{\label{subsec:Smax}Maximum entropy in nonequilibrium steady states}

We now extend the above principle of maximum entropy to 
nonequilibrium steady states without time evolution.
We consider specifically those cases where there is influx 
of current $J_{z}$ and/or energy
current $J_{\varepsilon z}$ through one boundary perpendicular to the $z$ axis.
It follows from the conservation laws that there is the same amount of 
currents flowing out through another boundary
in the steady state; 
hence these quantities can surely be adopted as 
additional variables of entropy to specify the system.

First of all, we assume that eq.\ (\ref{s_col>0}) also holds in steady states.
This can be proved explicitly within the second-order perturbation
as eq.\ (\ref{s_col^(2)}).
Next, we note that there is no net inflow or outflow of entropy
in steady states as well, i.e., the second term on the left-hand side
of eq.\ (\ref{s-flow-tot}) vanishes. 
Hence it is quite reasonable to expect that eq.\
(\ref{s>0}) also holds for steady states under appropriate conditions.
The question then arises: what are the parameters that have to be fixed?
In this context, we note that the principle of maximum entropy in
equilibrium can be stated without the magic word of ``isolated system''
as $S_{\rm eq}(E,V,N)\!=\!{\rm maximum}$, where energy $E$, volume $V$
and number $N$ are all mechanical variables.
This is quite natural, since ``probability'' can only be introduced at first
in terms of mechanical variables.
In contrast, temperature $T$, pressure $p$ and chemical potential $\mu$ are 
all equilibrium thermodynamic variables defined in terms of $S_{\rm eq}$
by partial differentiations, i.e.,
there are no definite definitions for them in nonequilibrium.
Hence the latter cannot specify nonequilibrium states of the system.
These considerations indicate that we should choose mechanical variables 
as independent variables of nonequilibrium entropy.
We hence add $J_{z}$ and $J_{\varepsilon z}$, which are conserved and
can be calculated mechanically,
as independent variables of entropy in the present context.
Now, we extend the principle of maximum entropy as follows:

{\em Principle of maximum entropy for steady states}:
The state which is realized most probably 
among possible steady states without time evolution
is the one that makes $S(E,V,N,J_{z},J_{\varepsilon z})$ maximum.

The validity of the principle can only be checked by
its consistency with experiments.
In the next paper we shall test it on Rayleigh-B\'enard convection
of a dilute classical gas which is typical of
nonequilibrium steady states with pattern 
formation.\cite{CH93,Chandrasekhar61,Busse78,Croquette89,
Koschmieder93,deBruyn96,BPA00}
It will be shown that the convection indeed 
gives rise to an increase of entropy over the value of the
heat conducting state.

Finally, it may be worth pointing out that,
once $S(E,V,N,J_{z},J_{\varepsilon z})$ is given explicitly,
we may perform successive Legendre transformations to 
change independent variables, just as in the equilibrium theory.

\section{Summary}

We have performed a theoretical study on entropy in 
nonequilibrium statistical mechanics
by specifically considering an assembly of identical bosons/fermions
interacting via a two-body potential.
First, we have presented an expression of nonequilibrium entropy density as
eq.\ (\ref{entropy-density}), which obeys the equation of motion (\ref{continuity-s}).
Thus, we can now trace time evolution of entropy in the
many-body system.
Second, we have proposed a principle of maximum entropy in \S\ref{subsec:Smax}
for nonequilibrium steady states without time evolution.
The validity of the principle will be checked in the next paper 
by calculating the entropy change of a dilute classical gas 
through the 
Rayleigh-B\'enard convective transition.

A conventional theoretical starting point to nonequilibrium systems has been some
phenomenological deterministic equations connected closely with 
the conservation laws.\cite{CH93}
One then performs the linear stability analysis
and derives some effective equations near the instability point
such as ``amplitude equations'' or ``phase equations.''
In some fortunate cases one may further be able to construct a Lyapunov function 
from those differential equations.\cite{CH93}
Note however that this approach is essentially of mechanical character,
as it is completely irrelevant to the concept of probability.
In contrast, little attention seems to have been paid to entropy in nonequilibrium 
systems, even near instability points, due partly to the absence of an explicit 
expression of nonequilibrium entropy.
Since entropy is the key concept of equilibrium thermodynamics and 
statistical mechanics embodying ``probability,'' it will be well worth studying
entropy of nonequilibrium systems and their ``phase transitions,''
which will shed new light on the phenomena.
The theoretical framework proposed here may provide a starting point for those investigations.
Its obvious advantage over the approach of the nonequilibrium statistical operator
by Zubarev\cite{Zubarev74} is that 
one can treat nonequilibrium systems which are globally
far away from equilibrium, though not locally.

Homogeneity/additivity has played a key role in constructing equilibrium statistical mechanics.
In contrast, the present approach may be regarded as an attempt
to treat open inhomogeneous systems
by an extremum principle with considering the boundary conditions explicitly.
Once it is established that steady states are identified correctly 
with the extremum principle, it will be a straightforward 
task to develop the linear-response
theory around it in the same way as in the equilibrium theory.\cite{Kubo57,Zubarev74}

\begin{acknowledgments}
The author would like to express his gratitude to H. R. Brand 
for useful and informative discussions
on nonequilibrium statistical mechanics and Rayleigh-B\'enard convection.
This work is supported in part by the 21st century COE program 
``Topological Science and  Technology,'' Hokkaido University.
\end{acknowledgments}

\appendix

\section{\label{App:conserve}
Conservation Laws}

We here show in terms of the present
real-time perturbation expansion in the Keldysh space
that various conservation laws are automatically satisfied
in the $\Phi$-derivative approximation.
We follow essentially the procedures by Baym\cite{Baym62}
with a slight modification appropriate for the real-time contour.

\subsection{Identities}

To start with, we derive several identities
which form the basis for proving the conservation laws.

First, consider the following gauge transformation:
\begin{subequations}
\label{gauge}
\begin{equation}
\check{G}(2,1)\longrightarrow 
{\rm e}^{i\check{\chi}(2)} 
\check{G}(2,1)
{\rm e}^{-i\check{\chi} (1)} \, ,
\label{gauge1}
\end{equation}
with
\begin{equation}
\check{\chi}(1)\equiv\left[
\begin{array}{cc}
\chi(1) & 0 \\
0 & 0
\end{array}
\right] .
\label{gauge2}
\end{equation}
\end{subequations}
It yields a first-order change in $\check{G}$ as
\begin{equation}
\delta\check{G}(2,1)= i\left[\check{\chi}(2) 
\check{G}(2,1)-
\check{G}(2,1)\check{\chi} (1)\right] \, .
\label{deltaG-gauge}
\end{equation}
However, $\Phi$ is clearly invariant through eq.\ (\ref{gauge}).
With eq.\ (\ref{Sigma-Phi}), this invariance of $\Phi$ reads
\begin{equation}
\int\! {\rm d}1\!\int\! {\rm d}2 \, {\rm Tr}
\check{\tau}_{3}\check{\Sigma}(1,2)\check{\tau}_{3} 
\delta \check{G}(2,1) = 0 \, .
\label{Identity}
\end{equation}
Substituting eq.\ (\ref{deltaG-gauge}) into eq.\ (\ref{Identity})
and using $\chi(1)$ is arbitrary, 
we obtain 
\begin{subequations}
\label{gauge-identity}
\begin{eqnarray}
&&\hspace{-6mm}
\int\! {\rm d}2 \,{\rm Tr}\, \frac{\check{1}\!+\!\check{\tau}_{3}}{2}
\bigl[ \check{\tau}_{3}\check{\Sigma}(1,2)\check{\tau}_{3} 
\check{G}(2,1)-
\check{G}(1,2)\check{\tau}_{3}
\check{\Sigma}(2,1)\check{\tau}_{3} 
\bigr] 
\nonumber \\
&&\hspace{-6mm}= 0 \, .
\label{gauge-identity1}
\end{eqnarray}
We can further transform eq.\ (\ref{gauge-identity1})
into
\begin{eqnarray}
&&\hspace{-15mm}
\int\! {\rm d}2\,
\bigl[ \Sigma^{{\rm R}}(1,2)G_{12}(2,1)
+ \Sigma_{12}(1,2) G^{{\rm A}}(2,1)
\nonumber \\
&&\hspace{-15mm}
-G^{{\rm R}}(1,2)\Sigma_{12}(2,1)
-G_{12}(1,2) \Sigma^{{\rm A}}(2,1)
 \bigr] =0\, ,
\label{gauge-identity2}
\end{eqnarray}
\end{subequations}
where we have used eqs.\ (\ref{G-prop2}), (\ref{Sigma-prop2}), (\ref{G-RAK-symm2}),
(\ref{Sigma-RAK-symm2})
and the Feynman rule (iv) around eq.\ (\ref{Feynman-rule-G}).
This is the basic identity obtained 
with respect to the gauge transformation.

Second, consider the following Galilean transformation:
\begin{subequations}
\label{Galilean}
\begin{equation}
\check{G}(2,1)\longrightarrow 
\exp\!\left[\check{\bm R}(t_{2})\!\cdot\!
\overrightarrow{\mbox{\boldmath$\nabla$}}_{2}\right]
\check{G}(2,1)
\exp\!\left[
\overleftarrow{\mbox{\boldmath$\nabla$}}_{1}\!\cdot\!\check{\bm R}(t_{1})\right] \, ,
\label{Galilean1}
\end{equation}
with
\begin{equation}
\check{\bm R}(t)\equiv \left[
\begin{array}{cc}
{\bm R}(t) & {\bm 0} \\
{\bm 0} & {\bm 0}
\end{array}
\right] .
\label{Galilean2}
\end{equation}
\end{subequations}
It yields the following first-order change:
\begin{equation}
\delta\check{G}(2,1)= 
\check{\bm R}(t_{2})\!\cdot\!{\mbox{\boldmath$\nabla$}}_{2}\check{G}(2,1) + 
\check{\bm R}(t_{1})\!\cdot\!{\mbox{\boldmath$\nabla$}}_{1}\check{G}(2,1) 
\, .
\label{deltaG-Galilean}
\end{equation}
However, $\Phi$ is invariant through eq.\ (\ref{Galilean}) so that
eq.\ (\ref{Identity}) holds also in this case.
Substituting (\ref{deltaG-Galilean}) into eq.\ (\ref{Identity}) and
using ${\bm R}(t)$ is arbitrary, we obtain
\begin{equation}
\!\int\!{\bm Q}(1)\, {\rm d}^{3}r_{1}={\bm 0} \, ,
\label{Galilean-identity}
\end{equation}
with ${\bm Q}(1)$ defined by
\begin{eqnarray}
&&\hspace{-11mm}
{\bm Q}(1)
\equiv \mp
i\hbar\frac{\mbox{\boldmath$\nabla$}_{\! 1}\!-\!\mbox{\boldmath$\nabla$}_{\! 1'}}{2}
\!\int\! {\rm d}2 \,
\bigl[
\Sigma^{\rm R}(1,2) G_{12}(2,1')
\nonumber \\
&&\hspace{1mm}
+\Sigma_{12}(1,2) G^{\rm A}(2,1')
- G^{\rm R}(1,2) \Sigma_{12}(2,1')
\nonumber \\
&&\hspace{1mm}
-G_{12}(1,2) \Sigma^{\rm A}(2,1')
\bigr]_{1'=1} 
\, .
\label{vecS}
\end{eqnarray}
Here terms with the self-energy derivative are due to
partial integrations.
Equation (\ref{Galilean-identity}) is the identity obtained 
in terms of the Galilean transformation.

We finally consider the change of time on $C_{1}$:
$t\! \rightarrow\! \theta(t)
\!\equiv\! t\!+\! \varphi(t)$.
Accordingly, $\check{G}$ is transformed as
\begin{subequations}
\label{rubber}
\begin{equation}
\check{G}(2,1)\longrightarrow \!
\check{U}(t_{2})
\check{G}({\bm r}_{2}\theta_{2},{\bm r}_{1}\theta_{1})\check{U}(t_{1})
\, ,
\label{rubber1}
\end{equation}
with
\begin{equation}
\check{U}(t)\equiv \left[
\begin{array}{cc}
({\rm d}\theta/{\rm d}t)^{1/4} & 0
\\
0 & 0
\end{array}
\right] .
\label{rubber2}
\end{equation}
\end{subequations}
The factor $({\rm d}\theta/{\rm d}t)^{1/4}$
cancels the Jacobian for 
$t\!\rightarrow\!\theta$ in eq.\ (\ref{S_C2}),
thereby keeping $\Phi$ invariant in form.
We hence conclude that eq.\ (\ref{Identity}) also holds in this case.
The first-order change in $\check{G}$ is given explicitly by
\begin{eqnarray}
&&\hspace{-14.5mm}
\delta G_{ji}(2,1)= 
\left\{\delta_{j1}\!\left[\frac{\varphi'(t_{2})}{4}
\!+\! \varphi(t_{2})\frac{\partial}{\partial t_{2}}\right]\right.
\nonumber \\
&&\hspace{4mm}
\left.+ \delta_{i1}\!\left[\frac{\varphi'(t_{1})}{4}
+ \varphi(t_{1})\frac{\partial}{\partial t_{1}}\right]\!\right\}
G_{ji}(2,1) \, .
\label{deltaG-rubber}
\end{eqnarray}
Substituting eq.\ (\ref{deltaG-rubber}) into eq.\
(\ref{Identity}) 
and using $\varphi(t)$ is arbitrary,
we obtain
\begin{equation}
\frac{{\rm d}\langle  H_{\rm int}(t_{1})\rangle}{{\rm d} t_{1}}
=-\int\!  Q_{\varepsilon}(1)\,{\rm d}^{3}r_{1}
\, ,
\label{rubber-identity}
\end{equation}
where $\langle  H_{\rm int}(t_{1})\rangle$ and
$Q_{\varepsilon}(1)$ are defined by
\begin{eqnarray}
&&\hspace{-10mm}
\langle H_{\rm int}(t_{1})\rangle=\pm
\frac{i\hbar}{4}\int\!{\rm d}^{3}r_{1}
\!\int \! {\rm d}2 \, \bigl[
\Sigma^{\rm R}(1,2) G_{12}(2,1)
\nonumber \\
&&\hspace{9.5mm}
+\Sigma_{12}(1,2) G^{\rm A}(2,1)
+ G^{\rm R}(1,2) \Sigma_{12}(2,1)
\nonumber \\
&&\hspace{9.5mm}
+G_{12}(1,2) \Sigma^{\rm A}(2,1)
\bigr]
\, ,
\label{Hint}
\end{eqnarray}
\begin{eqnarray}
&&\hspace{-10mm}
Q_{\varepsilon}(1)\equiv(\mp i\hbar) \frac{\partial}{\partial t_{1}}
\!\int \! {\rm d}2 \, \bigl[
\Sigma^{\rm R}(1',2) G_{12}(2,1)
\nonumber \\
&&\hspace{4mm}
+\Sigma_{12}(1',2) G^{\rm A}(2,1)+ G^{\rm R}(1,2) \Sigma_{12}(2,1')
\nonumber \\
&&\hspace{4mm}
+G_{12}(1,2) \Sigma^{\rm A}(2,1')
\bigr]_{1'=1} \, ,
\label{Q}
\end{eqnarray}
respectively.
Equation (\ref{rubber-identity}) is the identity obtained from eq.\ (\ref{rubber}).

It should be noted that $\langle  H_{\rm int}(t_{1})\rangle$ defined above is 
exactly the interaction energy of the system.
This can be seen as follows:
With eqs.\ (\ref{G-prop}),
(\ref{Sigma-prop}), (\ref{G-RAK-symm2}) and (\ref{Sigma-RAK-symm2}), 
the $12$ element of eq.\ (\ref{Dyson-check})
is transformed into
\begin{eqnarray}
&&\hspace{-4mm}
\left[i\hbar \frac{\partial }{\partial t_{1}}\!+\! 
\frac{\hbar^{2}\nabla_{1}^{2}}{2m}\!-\!U(1)
\right]\!
G_{12}(1,2)-
\int \!\left[\Sigma^{{\rm R}}(1,3)G_{12}(3,2)\right.
\nonumber \\
&&\hspace{-4mm}
\left.
+\,
\Sigma_{12}(1,3)G^{{\rm A}}(3,2)\right]{\rm d}3 =0 \, .
\label{Dyson-12-l}
\end{eqnarray}
We may alternatively derive an equation for 
$G_{12}(1,2)\!=\!\mp(i/\hbar)\langle\psi_{\cal H}^{\dagger}(2)
\psi_{\cal H}(1)\rangle$ starting from the equation of motion 
for $\psi_{\cal H}(1)$:
\begin{eqnarray}
&&\hspace{-10mm}
\!\left[i\hbar \frac{\partial }{\partial t_{1}}+
\frac{\hbar^{2}\nabla_{1}^{2}}{2m}-U(1)
\right]\! \psi_{\cal H}(1)
\nonumber \\
&&\hspace{-10mm}
- \int {\rm d}1'\,
\bar{V}(1\!-\!1') \psi_{\cal H}^{\dagger}(1')\psi_{\cal H}(1')
\psi_{\cal H}(1) =0 \, .
\label{eq-motion}
\end{eqnarray}
Let us multiply eq.\ (\ref{eq-motion}) by $(\mp i/\hbar)\psi_{\cal H}^{\dagger}(2)$
from the left,
take its thermodynamic average, and compare the result with
eq.\ (\ref{Dyson-12-l}).
We then obtain the identity:
\begin{eqnarray}
&&\hspace{-10mm}
\int {\rm d}1'\,
\bar{V}(1\!-\!1') \langle\psi_{\cal H}^{\dagger}(2)\psi_{\cal H}^{\dagger}(1')
\psi_{\cal H}(1')
\psi_{\cal H}(1)\rangle
\nonumber \\
&&\hspace{-10mm}
=\pm i\hbar \int \!\!\left[\Sigma^{{\rm R}}(1,3)G_{12}(3,2)
\!+\! \Sigma_{12}(1,3)G^{{\rm A}}(3,2)\right]{\rm d}3 \, .
\nonumber \\
&&
\label{Hint-identity}
\end{eqnarray}
Setting $2\!=\! 1$ in eq.\ (\ref{Hint-identity})
yields an expression of the interaction energy in terms of the self-energy.
An alternative expression is obtained by taking its complex conjugate.
We thereby conclude that eq.\ (\ref{Hint}) is indeed 
the interaction energy of the system.

\subsection{Conservation laws}
\label{subsec:conserve}

We are now ready to prove that the conservation laws are 
automatically satisfied in
the $\Phi$-derivative approximation.

Let us take complex conjugate of eq.\ (\ref{Dyson-12-l})
and use eqs.\ (\ref{G-prop1}), (\ref{checkSigma-prop}), (\ref{G-RAK-symm1}) and
(\ref{Sigma-RAK-symm1}). This yields
\begin{eqnarray}
&&\hspace{-4mm}
\left[-i\hbar \frac{\partial }{\partial t_{2}}\!+\! 
\frac{\hbar^{2}\nabla_{2}^{2}}{2m}\!-\!U(2)
\right]\!
G_{12}(1,2)-\!
\int \!\!\left[G^{{\rm R}}(1,3)\Sigma_{12}(3,2)\!\right.
\nonumber \\
&&\hspace{-4mm}
\left.
+\,G_{12}(1,3)\Sigma^{{\rm A}}(3,2)\right]{\rm d}3 =0 \, .
\label{Dyson-12-r}
\end{eqnarray}
First, we subtract eq.\ (\ref{Dyson-12-r}) from eq.\ (\ref{Dyson-12-l}),
set $2\!=\!1$, and use the identity (\ref{gauge-identity2}).
We thereby obtain the number conservation law as
\begin{equation}
\frac{\partial n(1)}{\partial t_{1}}+{\bm\nabla}_{1}\!\cdot{\bm j}(1)=0 \, ,
\label{conserve-n}
\end{equation}
where $n(1)$ and ${\bm j}(1)$ are the particle and current densities
defined by
\begin{subequations}
\label{nj}
\begin{equation}
n(1)\equiv \pm i\hbar G_{12}(1,1) \, ,
\label{number}
\end{equation}
\begin{equation}
{\bm j}(1)\equiv \left.
\pm \hbar^{2}\frac{{\bm\nabla}_{1}\!-\!{\bm\nabla}_{2}}{2m} G_{12}(1,2) 
\right|_{2=1}\, ,
\label{current}
\end{equation}
\end{subequations}
respectively.

We next operate $\mp i\hbar({\bm\nabla}_{1}\!-\!{\bm\nabla}_{2})/2m$ to eqs.\
(\ref{Dyson-12-l}) and (\ref{Dyson-12-r}), subtract the
latter from the former, and set $2\!=\! 1$.
We thereby find that the time evolution of the current density
obeys
\begin{equation}
\frac{\partial}{\partial t_{1}}{\bm j}(1)
+\frac{1}{m}{\bm\nabla}_{1}\underline{\Theta}^{K}(1)
+\frac{n(1)}{m}{\bm\nabla}_{1}U(1)
=\frac{1}{m}{\bm Q}(1) \, ,
\label{momentum-eq}
\end{equation}
where ${\bm Q}(1)$ is given by eq.\ (\ref{vecS}),
and tensor $\underline{\Theta}^{K}(1)$ is defined by
\begin{equation}
\Theta_{ij}^{K}(1)=\mp \frac{i\hbar^{3}}{4m}
(\nabla_{\! 1i}\!-\!\nabla_{\! 2i})
(\nabla_{\! 1j}\!-\!\nabla_{\! 2j})G_{12}(1,2)\bigr|_{2=1} \, .
\label{Pi^K}
\end{equation}
Integrating eq.\ (\ref{momentum-eq}) over the whole space of the relevant system
and using eq.\ (\ref{Galilean-identity}), we obtain the total momentum conservation
law as
\begin{equation}
\frac{\partial}{\partial t_{1}}\!\int\! {\bm j}(1)\,{\rm d}^{3}r_{1}
=-\!\int\! \frac{n(1)}{m}{\bm\nabla}_{1}U(1)\,{\rm d}^{3}r_{1}\, .
\label{conserve-p}
\end{equation}

We finally operate $\mp i\hbar\frac{\partial}{\partial t_{2}}$
and $\mp i\hbar\frac{\partial}{\partial t_{1}}$ to eqs.\ 
(\ref{Dyson-12-l}) and (\ref{Dyson-12-r}), respectively.
Adding the resulting equations and setting $2\!=\! 1$,
we obtain
\begin{equation}
\frac{\partial {\cal E}_{{\rm K}}(1)}{\partial t_{1}} 
+{\bm\nabla}_{1}\cdot{\bm j}_{\varepsilon}'(1)
+U(1)\frac{\partial n(1)}{\partial t_{1}}
=Q_{\varepsilon}(1) \, ,
\label{energy-eq}
\end{equation}
where $Q_{\varepsilon}(1)$ is given by eq.\ (\ref{Q}),
and ${\cal E}_{{\rm K}}(1)$ and ${\bm j}_{\varepsilon}'(1)$ are defined by
\begin{equation}
{\cal E}_{{\rm K}}(1)\equiv \pm  \frac{i\hbar^{3}}{2m}{\bm \nabla}_{1}\!\cdot\!{\bm\nabla}_{2}
G_{12}(1,2)\biggr|_{2=1} \, ,
\label{Kinetic-e}
\end{equation}
\begin{equation}
{\bm j}_{\varepsilon}'(1)=\mp \frac{i\hbar^{3}}{2m}
\!\left(\frac{\partial}{\partial t_{1}}{\bm\nabla}_{2}\!+\!
\frac{\partial}{\partial t_{2}}{\bm\nabla}_{1}\right)\! G_{12}(1,2)\biggr|_{2=1}\, ,
\label{j_e'}
\end{equation}
respectively.
The quantity ${\cal E}_{{\rm K}}(1)$ denotes the kinetic energy density.
We next integrate eq.\ (\ref{energy-eq}) over space and use eqs.\ 
(\ref{rubber-identity}) and (\ref{conserve-n}).
We thereby obtain the total-energy conservation law as
\begin{equation}
\frac{{\rm d}}{{\rm d} t_{1}}\!\left[\int\! {\cal E}_{{\rm K}}(1)\,{\rm d}^{3}r_{1}
\!+\!\langle H_{\rm int}(t_{1})\rangle\right]\!=-\!
\int \!{\bm j}(1)\cdot\!{\bm\nabla}_{1}U(1)\,{\rm d}^{3}r_{1}
\, .
\label{conserve-e}
\end{equation}

Thus, the conservation laws are automatically
satisfied in the $\Phi$-derivative
approximation as eqs.\ (\ref{conserve-n}), (\ref{conserve-p})
and (\ref{conserve-e}).

\section{\label{App:Gamma}
Expression of $\Gamma$}

We here present expressions of the vertex function (\ref{Xi})
within the second-order skeleton expansion by
using eqs.\ (\ref{Phi^(1)}) and (\ref{Phi^(2)}).
This will help us to understand the structures of $\Gamma$.

First, $\Gamma^{(1)}_{ii',jj'}$ is obtained from eq.\ (\ref{Phi^(1)})
with eq.\ (\ref{Xi}).
It may be written compactly as
\begin{eqnarray}
&&\hspace{-5mm}
\Gamma^{(1)}_{ii',jj'}(11',22')
= \bar{V}(1\!-\!2)
\delta_{ij}\delta_{i'j'}(\check{\tau}_{3})_{ii'}[
\delta(1,1')\delta(2,2')
\nonumber \\
&&\hspace{23mm}
\pm\delta(1,2')\delta(2,1')]\, .
\label{Lambda^(1)_ii'jj'}
\end{eqnarray}
Thus, only $\Gamma^{(1)}_{11,11}=-
\Gamma^{(1)}_{22,22}$ are finite among $\Gamma^{(1)}_{ii',jj'}$.

Second, $\Gamma^{(2)}_{ii',jj'}$ is calculated
from eq.\ (\ref{Phi^(2)}). 
It turns out that only $\Gamma^{(2)}_{ij,ij}$ and
$\Gamma^{(2)}_{ij,ji}$ are finite.
For $i\!\neq\!j$, they are given by
\begin{subequations}
\begin{eqnarray}
&&\hspace{-10mm}
\Gamma^{(2)}_{ij,ij}(11',22')
=-i\hbar \bar{V}(1\!-\!2)\bar{V}(1'\!-\!2')[G_{ji}(1',1)
\nonumber \\
&&\hspace{0mm}
\times G_{ji}(2',2) \pm G_{ji}(1',2)G_{ji}(2',1)]  \, ,
\label{Lambda^(2)_2121}
\end{eqnarray}
\begin{eqnarray}
&&\hspace{-7mm}
\Gamma^{(2)}_{ij,ji}(11',22')
\nonumber \\
&&\hspace{-7mm}
=-i\hbar \biggl[ \bar{V}(1'\!-\!2)\bar{V}(1\!-\!2')G_{ji}(1',1)G_{ij}(2',2)
\nonumber \\
&&\hspace{-7mm}
+\delta(1',2)\delta(1,2')\!\int\! {\rm d}3\!\int\!{\rm d}3'\,
\bar{V}(1'\!-\!3)\bar{V}(1\!-\!3')
\nonumber \\
&&\hspace{20mm}\times G_{ji}(3,3')G_{ij}(3',3)
\nonumber \\
&&\hspace{-7mm}
\pm \delta(1',2)\bar{V}(1\!-\!2')
\!\int\! {\rm d}3\,\bar{V}(1'\!-\!3)G_{ij}(2',3)G_{ji}(3,1)
\nonumber \\
&&\hspace{-7mm}
\pm \bar{V}(1'\!-\!2)\delta(1,2')
\!\int\! {\rm d}3\,\bar{V}(1\!-\!3)G_{ji}(1',3)G_{ij}(3,2) \biggr]
\, ,
\nonumber \\
&&
\label{Lambda^(2)_1221}
\end{eqnarray}
respectively. Also, $\Gamma^{(2)}_{ii,ii}$ is given in terms of 
the above quantities as
\begin{eqnarray}
&&\hspace{-10mm}
\Gamma^{(2)}_{11,11}(11',22')=-\Gamma^{(2)*}_{22,22}(1'1,2'2)
\nonumber \\
&&\hspace{-13.7mm}
=-\theta(t_{1}'-t_{1})\theta(t_{2}'-t_{2})\Gamma^{(2)}_{12,12}(11',22')
\nonumber \\
&&\hspace{-10mm}
-\theta(t_{1}'-t_{1})\theta(t_{2}-t_{2}')\Gamma^{(2)}_{12,21}(11',22')
\nonumber \\
&&\hspace{-10mm}
-\theta(t_{1}-t_{1}')\theta(t_{2}-t_{2}')\Gamma^{(2)}_{21,21}(11',22')
\nonumber \\
&&\hspace{-10mm}
-\theta(t_{1}-t_{1}')\theta(t_{2}'-t_{2})\Gamma^{(2)}_{21,12}(11',22')\, .
\label{Lambda^(2)_1111}
\end{eqnarray}
\end{subequations}
These are the expressions 
of $\Gamma^{(n)}_{ii',jj'}$ for $n\!=\!1,2$ in the space-time coordinates.

We next write down $\Gamma_{ii',jj'}(11',22')$ of the local approximation
which will be necessary later.
In this case, $\Gamma_{ii',jj'}(11',22')$
acquires the structure of the uniform system.
To be specific, we substitute eq.\ (\ref{checkG-Wig}) into eq.\ (\ref{Xi})
and adopt the local approximation.
We then find order by order that $\Gamma_{ii',jj'}(11',22')$ 
can be expanded as
\begin{eqnarray}
&&\hspace{-6mm}
\Gamma_{ii',jj'}(11',22')
\nonumber \\ 
&&\hspace{-6mm}=\int\frac{{\rm d}^{3}p_{1}{\rm d}\varepsilon_{1}}{(2\pi\hbar)^{4}}
\int\frac{{\rm d}^{3}p_{2}{\rm d}\varepsilon_{2}}{(2\pi\hbar)^{4}}
\int\frac{{\rm d}^{3}q\,{\rm d}\omega}{(2\pi\hbar)^{4}}
{\rm e}^{-i({\bm p}_{1+}\cdot{\bm r}_{1}-\varepsilon_{1+} t_{1})/\hbar}
\nonumber \\
&&\hspace{-2mm} 
\times{\rm e}^{i[({\bm p}_{1-}\cdot{\bm r}_{1}'-\varepsilon_{1-} t_{1}')
-({\bm p}_{2-}\cdot{\bm r}_{2}-\varepsilon_{2-} t_{2})
+({\bm p}_{2+}\cdot{\bm r}_{2}'-\varepsilon_{2+} t_{2}')]/\hbar}
\nonumber \\
&&\hspace{-2mm} 
\times \Gamma_{ii',jj'}({\bm p}_{1}\varepsilon_{1},{\bm p}_{2}\varepsilon_{2};
{\bm q}\omega,{\bm r}t) \, ,
\label{Xi-Wig}
\end{eqnarray}
where ${\bm p}_{j\pm}\!\equiv\!{\bm p}_{j}\pm{\bm q}/2$ and
$\varepsilon_{j\pm}\!\equiv\!\varepsilon_{j}\pm\omega/2$
in this expression, and ${\bm r}t$
denotes a local space-time point around $1$, $1'$, $2$ and $2'$.
Let us write down $\Gamma^{(n)}_{ii',jj'}(p_{1},p_{2},q)$
explicitly for $n\!=\! 1,2$ with 
$p\!\equiv\!{\bm p}\varepsilon$ and $q\!\equiv\!{\bm q}\omega$.
First, $\Gamma^{(1)}_{11,11}(p_{1},p_{2},q)$
is obtained from eq.\ (\ref{Lambda^(1)_ii'jj'}) as
\begin{equation}
\Gamma^{(1)}_{11,11}(p_{1},p_{2},q)=
V_{\bm q}\pm V_{{\bm p}_{1}-{\bm p}_{2}} \, .
\end{equation}
Equations (\ref{Lambda^(2)_2121}) and (\ref{Lambda^(2)_1221}) for $i\!\neq\!j$
are transformed similarly into
\begin{subequations}
\begin{eqnarray}
&&\hspace{-4mm}
\Gamma^{(2)}_{ij,ij}(p_{1},p_{2},q)
=-i\hbar\int\!\frac{{\rm d}^{4}q'}{(2\pi\hbar)^{4}}
G_{ji}(p_{1}\!-\!q')G_{ji}(p_{2}\!+\!q')
\nonumber \\
&&\hspace{4mm}
\times V_{{\bm q}'+{\bm q}/2}\bigl(V_{{\bm q}'-{\bm q}/2}\!\pm \! 
V_{{\bm p}_{2}-{\bm p}_{1}+{\bm q}'+{\bm q}/2}\bigr)
\, ,
\end{eqnarray}
\begin{eqnarray}
&&\hspace{-4mm}
\Gamma^{(2)}_{ij,ji}(p_{1},p_{2},q)
=-i\hbar\int\frac{{\rm d}^{4}q'}{(2\pi\hbar)^{4}}
G_{ji}(p_{1}\!+\!q')G_{ij}(p_{2}\!+\!q')
\nonumber \\
&&\hspace{4mm}
\times
\bigl(V_{{\bm q}'-{\bm q}/2}\!\pm\!V_{{\bm p}_{1}-{\bm p}_{2}}\bigr)
\bigl(V_{{\bm q}'+{\bm q}/2}\!\pm\!V_{{\bm p}_{1}-{\bm p}_{2}}\bigr)
\, .
\end{eqnarray}
Also, eq.\ (\ref{Lambda^(2)_1111}) yields
\begin{eqnarray}
&&\hspace{-4mm}
\Gamma^{(2)}_{11,11}(p_{1},p_{2},q)=
\int\frac{{\rm d}\varepsilon_{1}'}{2\pi}
\int\frac{{\rm d}\varepsilon_{2}'}{2\pi}\biggl[
\frac{\Gamma^{(2)}_{12,12}({\bm p}_{1}\varepsilon_{1}',{\bm p}_{2}\varepsilon_{2}',
{\bm q}\omega)}{(\varepsilon_{1+}\!-\!\varepsilon_{1}')
(\varepsilon_{2+}\!-\!\varepsilon_{2}')}
\nonumber \\
&&\hspace{6mm}
-\frac{\Gamma^{(2)}_{12,21}({\bm p}_{1}\varepsilon_{1}',{\bm p}_{2}\varepsilon_{2}',
{\bm q}\omega)}{(\varepsilon_{1+}\!-\!\varepsilon_{1}')
(\varepsilon_{2-}\!-\!\varepsilon_{2}')}
+\frac{\Gamma^{(2)}_{21,21}({\bm p}_{1}\varepsilon_{1}',{\bm p}_{2}\varepsilon_{2}',
{\bm q}\omega)}{(\varepsilon_{1-}\!-\!\varepsilon_{1}')
(\varepsilon_{2-}\!-\!\varepsilon_{2}')}
\nonumber \\
&&\hspace{6mm}
-\frac{\Gamma^{(2)}_{21,12}({\bm p}_{1}\varepsilon_{1}',{\bm p}_{2}\varepsilon_{2}',
{\bm q}\omega)}{(\varepsilon_{1-}\!-\!\varepsilon_{1}')
(\varepsilon_{2+}\!-\!\varepsilon_{2}')}\biggr] \, ,
\end{eqnarray}
\end{subequations}
with $\varepsilon_{j\pm}\!\equiv\!\varepsilon_{j}\!\pm\!i0_{+}$ in this expression.

\section{\label{app:conserve-W}Conservation laws in the Wigner representation}

Equations (\ref{conserve-n}), (\ref{momentum-eq}) and (\ref{energy-eq})
have a fundamental importance of describing the flows of particle,
momentum and energy.
We here transform these differential conservation laws into the Wigner representation
within the first-order gradient expansion.

Substituting eqs.\ (\ref{checkG-Wig}) and (\ref{G_12-Wig}) into eq.\ (\ref{nj}), 
we can write $n({\bm r}t)$ and ${\bm j}({\bm r}t)$ alternatively in terms of 
$A$ and $\phi$. 
Those expressions are formally exact and satisfy eq.\ (\ref{conserve-n}).
The density $n({\bm r}t)$ and the local velocity $
{\bm v}({\bm r}t)\!\equiv\!{\bm j}({\bm r}t)/n({\bm r}t)$
are now given by
\begin{subequations}
\label{nj-p}
\begin{equation}
n({\bm r}t)= 
\hbar\int\!\frac{{\rm d}^{3}p\,{\rm d}\varepsilon}{(2\pi\hbar)^{4}}
A({\bm p}\varepsilon,{\bm r}t)\phi({\bm p}\varepsilon,{\bm r}t) \, ,
\label{rho}
\end{equation}
\begin{equation}
{\bm v}({\bm r}t)
=\frac{\hbar}{n({\bm r}t)}\int\!\frac{{\rm d}^{3}p\,{\rm d}\varepsilon}
{(2\pi\hbar)^{4}}\,
\frac{\bm p}{m}A({\bm p}\varepsilon,{\bm r}t)\phi({\bm p}\varepsilon,{\bm r}t) \, ,
\label{v}
\end{equation}
\end{subequations}
respectively.
The particle conservation law (\ref{conserve-n}) then reads
\begin{equation}
\frac{\partial n}{\partial t}+{\bm \nabla}(n{\bm v})
=0 \, .
\label{continuity}
\end{equation}

We next consider eq.\ (\ref{momentum-eq}) for the momentum flow.
Here it is desirable to express
${\bm Q}(1)$ on the right-hand side as a divergence.
To carry this out within the first-order gradient expansion,
we use eq.\ (\ref{Hint-identity}) and its complex conjugate in
eq.\ (\ref{vecS}).
We then obtain an alternative expression of ${\bm Q}(1)$ as
$$
{\bm Q}(1)\!
=\!-\!\int\!{\rm d}^{3}r_{1}'
\frac{\partial V({\bm r}_{1}\!-\!{\bm r}_{1}')}{\partial {\bm r}_{1}}
\langle\psi_{\cal H}^{\dagger}(1)\psi_{\cal H}^{\dagger}(1')
\psi_{\cal H}(1')
\psi_{\cal H}(1)\rangle ,
$$
with $1'\!=\!{\bm r}_{1}'t_{1}$ in this expression.
We further write
$\langle\psi_{\cal H}^{\dagger}(1)\psi_{\cal H}^{\dagger}(1')
\psi_{\cal H}(1')
\psi_{\cal H}(1)\rangle\!=\!\rho_{2}\bigl({\bm r}_{1}\!-\!{\bm r}_{1}',
\frac{{\bm r}_{1}+{\bm r}_{1}'}{2},t_{1}\bigr)$ and expand the argument
$\frac{{\bm r}_{1}+{\bm r}_{1}'}{2}$ from ${\bm r}_{1}$
up to the first order.\cite{KB62,Zubarev74} 
We thereby obtain
\begin{equation}
{\bm Q}(1) = -{\bm \nabla}_{1} \underline{\Pi}^{V}(1)
\, ,
\label{vecQ-2}
\end{equation}
where tensor $\underline{\Pi}^{V}$ is defined by
\begin{eqnarray}
&&\hspace{-6mm}
\Pi_{ij}^{V}(1)\equiv-\frac{1}{2}
\int{\rm d}^{3}\bar{r}\,
\bar{r}_{i} \frac{\partial V(\bar{\bm r})}{\partial \bar{r}_{j}}
\langle\psi_{\cal H}^{\dagger}(1_{+})\psi_{\cal H}^{\dagger}(1_{-})
\psi_{\cal H}(1_{-})
\nonumber \\
&&\hspace{7mm}
\times
\psi_{\cal H}(1_{+})\rangle \, ,
\label{Pi^V}
\end{eqnarray}
with $1_{\pm}\!\equiv\!({\bm r}_{1}\!\pm\!{\bar{\bm r}}/{2},t_{1})$.
Let us substitute eq.\ (\ref{vecQ-2}) into eq.\ (\ref{momentum-eq}).
We then arrive at the differential momentum conservation law as
\begin{equation}
\frac{\partial}{\partial t}{\bm j}({\bm r}t)
+\frac{1}{m}{\bm\nabla}\underline{\Theta}({\bm r}t)
=-\frac{n({\bm r}t)}{m}{\bm\nabla}U({\bm r}t) \, ,
\label{momentum-eq-grad}
\end{equation}
where tensor $\Theta_{ij}$ is defined by
\begin{equation}
\Theta_{ij}({\bm r}t)\equiv \Theta_{ij}^{K}({\bm r}t) 
+\Pi_{ij}^{V}({\bm r}t) \, .
\label{Pi}
\end{equation}
The quantity $\Theta_{ij}^{K}$, given by eq.\ (\ref{Pi^K}),
may be written alternatively in terms of $A$ and $f$ as
\begin{eqnarray}
&&\hspace{-10mm}
\Theta_{ij}^{K}({\bm r}t)= \hbar
\!\int\! \frac{{\rm d}^{3}p\,{\rm d}\varepsilon}{(2\pi\hbar)^{4}}\,
\frac{p_{i}p_{j}}{m}
A({\bm p}\varepsilon,{\bm r}t)
\phi({\bm p}\varepsilon,{\bm r}t) 
\nonumber \\
&&\hspace{2.5mm}
= mn({\bm r}t) v_{i}({\bm r}t)v_{j}({\bm r}t) +\Pi_{ij}^{K}({\bm r}t) \, ,
\label{tildeTheta^K-p}
\end{eqnarray}
where
\begin{equation}
\Pi_{ij}^{K}({\bm r}t) \equiv \hbar
\!\int\! \frac{{\rm d}^{3}p\,{\rm d}\varepsilon}{(2\pi\hbar)^{4}}\,
\frac{\bar{p}_{i}\bar{p}_{j}}{m}
A({\bm p}\varepsilon,{\bm r}t)
\phi({\bm p}\varepsilon,{\bm r}t) \, ,
\label{tildePi^K}
\end{equation}
with $\bar{\bm p}\!\equiv\!{\bm p}-m{\bm v}$.
This $\Pi_{ij}^{K}({\bm r}t)$ denotes kinetic part of
the momentum flux density tensor 
in the coordinate system moving with the local velocity ${\bm v}({\bm r}t)$.
Using eqs.\ (\ref{nj-p}), (\ref{continuity}), (\ref{Pi}) and (\ref{tildeTheta^K-p}),
we can transform eq.\ (\ref{momentum-eq-grad}) further into
\begin{equation}
\frac{\partial {\bm v}}{\partial t}+{\bm v}\!\cdot\!{\bm \nabla}{\bm v}
+\frac{1}{mn}{\bm \nabla}\underline{\Pi}=-\frac{{\bm\nabla}U}{m} \, ,
\label{NS}
\end{equation}
where tensor $\Pi_{ij}$ is defined by
\begin{equation}
\Pi_{ij}({\bm r}t)\equiv \Pi_{ij}^{K}({\bm r}t) 
+\Pi_{ij}^{V}({\bm r}t) \, ,
\label{tildePi}
\end{equation}
with $\Pi_{ij}^{K}$ and $\Pi_{ij}^{V}$ given by eqs.\
(\ref{tildePi^K}) and (\ref{Pi^V}), respectively.

It remains to evaluate eq.\ (\ref{Pi^V}) within
the first-order gradient expansion of the $\Phi$-derivative approximation.
This is essentially the calculation of the two-particle correlation function
${\cal K}$ defined by eq.\ (\ref{K}). Since $\Pi_{ij}^{V}({\bm r}t)$ 
is operated by ${\bm \nabla}$ in eq.\ (\ref{momentum-eq-grad}), 
we can adopt the local approximation for this purpose.
Thus, the procedure to obtain ${\cal K}_{ii',jj'}(11',22')$ 
is exactly the same as that for the uniform system.
To be specific, let us substitute 
eqs.\ (\ref{checkG-Wig}) and (\ref{Xi-Wig}) into eq.\ (\ref{K-2}).
We then find that ${\cal K}$ can also be expanded as
\begin{eqnarray}
&&\hspace{-6mm}
{\cal K}_{ii',jj'}(11',22')
\nonumber \\ 
&&\hspace{-6mm}=\int\frac{{\rm d}^{3}p_{1}{\rm d}\varepsilon_{1}}{(2\pi\hbar)^{4}}
\int\frac{{\rm d}^{3}p_{2}{\rm d}\varepsilon_{2}}{(2\pi\hbar)^{4}}
\int\frac{{\rm d}^{3}q\,{\rm d}\omega}{(2\pi\hbar)^{4}}
{\rm e}^{i({\bm p}_{1-}\cdot{\bm r}_{1}-\varepsilon_{1-} t_{1})/\hbar}
\nonumber \\
&&\hspace{-2mm} 
\times{\rm e}^{i[-({\bm p}_{1+}\cdot{\bm r}_{1}'-\varepsilon_{1+} t_{1}')
+({\bm p}_{2+}\cdot{\bm r}_{2}-\varepsilon_{2+} t_{2})
-({\bm p}_{2-}\cdot{\bm r}_{2}'-\varepsilon_{2-} t_{2}')]/\hbar}
\nonumber \\
&&\hspace{-2mm} 
\times {\cal K}_{ii',jj'}({\bm p}_{1}\varepsilon_{1},{\bm p}_{2}\varepsilon_{2};
{\bm q}\omega,{\bm r}t) \, .
\label{K-Wig}
\end{eqnarray}
The quantity ${\cal K}_{ii',jj'}
({\bm p}\varepsilon,{\bm p}'\varepsilon';{\bm q}\omega,
{\bm r}t)$ satisfies
\begin{equation}
\check{\check{\mbox{$\cal K$}}}=\pm \bigl(\,\check{\check{1}}\mp i\hbar\,
\check{G}\check{G}\,\check{\check{\Gamma}}\,\bigr)^{-1}\check{G}\check{G} \, ,
\label{K-2p}
\end{equation}
where every quantity should be regarded now as a matrix
in terms of ${\bm p}\varepsilon$-${\bm p}'\varepsilon'$ instead of 
$11'$-$22'$ in eq.\ (\ref{K-2}), with integration
$\int {\rm d}^{3}p\,{\rm d}\varepsilon/(2\pi\hbar)^{4}$ over every
internal variable ${\bm p}\varepsilon$ implied.
Indeed, $\check{\check{1}}$ and $\check{G}\check{G}$ are now defined by
\begin{subequations}
\label{checkcheck-p}
\begin{equation}
(\check{\check{1}})_{ii',jj'}({\bm p}\varepsilon,{\bm p}'\varepsilon')
=\delta_{ij}\delta_{i'j'}
(2\pi\hbar)^{4}\delta({\bm p}\!-\!{\bm p}')\delta(\varepsilon\!-\!\varepsilon')\, ,
\end{equation}
\begin{eqnarray}
&&\hspace{-4mm}
(\check{G}\check{G})_{ii',jj'}({\bm p}\varepsilon,{\bm p}'\varepsilon';
{\bm q}\omega,{\bm r}t)
\nonumber \\
&&\hspace{-4mm}
=G_{ij'}({\bm p}_{-}\varepsilon_{-},{\bm r}t)G_{ji'}({\bm p}_{+}\varepsilon_{+},{\bm r}t)
(2\pi\hbar)^{4}\delta({\bm p}\!-\!{\bm p}')\delta(\varepsilon\!-\!\varepsilon') \, ,
\nonumber \\
&&
\label{GG-p}
\end{eqnarray}
\end{subequations}
respectively,
with ${\bm p}_{\pm}\!\equiv\!{\bm p}\pm{\bm q}/2$ and
$\varepsilon_{\pm}\!\equiv\!\varepsilon\pm\omega/2$ in this expression.
Equation (\ref{K-2p}) enables us to calculate two-particle correlation
functions for a given vertex function $\Gamma$, which in turn
is specified completely for a given $\Phi$. 

We now express $\Pi^{V}_{ij}$ of eq.\ (\ref{Pi^V})
in terms of the solution of eq.\ (\ref{K-2p}).
We first rewrite eq.\ (\ref{Pi^V}) by using
${\cal K}$ of eq.\ (\ref{K}).
We then substitute eqs.\ (\ref{V-Fourier}), (\ref{checkG-Wig}) and (\ref{K-Wig}) 
into eq.\ (\ref{Pi^V}), remove $\bar{r}_{i}$ by using 
$\bar{r}_{i}\,{\rm e}^{i{\bm q}\cdot\bar{\bm r}_{}/\hbar}
=-i\hbar\frac{\partial}{\partial q_{i}}
{\rm e}^{i{\bm q}\cdot\bar{\bm r}/\hbar}$,
perform partial integrations over ${\bm q}$,
and carry out the integration over 
$\bar{\bm r}$.
We thereby arrive at an alternative expression of tensor $\Pi^{V}_{ij}$.
Substituting it as well as eq.\ (\ref{tildePi^K}) into
eq.\ (\ref{tildePi}), we obtain
\begin{eqnarray}
&&\hspace{-12mm}
\Pi_{ij}({\bm r}t)=\hbar
\int \frac{{\rm d}^{3}p\,{\rm d}\varepsilon}{(2\pi\hbar)^{4}}\,
\frac{\bar{p}_{i}\bar{p}_{j}}{m}
A({\bm p}\varepsilon,{\bm r}t)
\phi({\bm p}\varepsilon,{\bm r}t)
\nonumber \\
&&\hspace{-7mm}
+\frac{(i\hbar)^{2}}{2}\int 
\frac{{\rm d}^{3}q\,{\rm d}\omega}{(2\pi\hbar)^{4}}
\int \frac{{\rm d}^{3}p\,{\rm d}\varepsilon}{(2\pi\hbar)^{4}}
\int \frac{{\rm d}^{3}p'{\rm d}\varepsilon'}{(2\pi\hbar)^{4}}
\nonumber \\
&&\hspace{-7mm}
\times 
\!\left(V_{q}\delta_{ij}\!+\!\frac{q_{i}q_{j}}{q}\frac{{\rm d}V_{q}}{{\rm d}q}\right)\!
\bigl[{\cal K}_{12,12}({\bm p}\varepsilon,{\bm p}'\varepsilon';{\bm q}\omega,{\bm r}t)
\nonumber \\
&&\hspace{-7mm}
+(2\pi\hbar)^{4}
\delta({\bm q})\delta(\omega)G_{12}({\bm p}\varepsilon,{\bm r}t)
G_{12}({\bm p}'\varepsilon',{\bm r}t)
\bigr] \, ,
\label{tildePi2}
\end{eqnarray}
where $\bar{\bm p}\!=\!{\bm p}\!-\!m{\bm v}$, 
and $G_{12}$ and ${\cal K}$ are given by eqs.\ (\ref{G_12-Wig}) and (\ref{K-2p}),
respectively.
We observe clearly that $\underline{\Pi}$ is a symmetric tensor.

We finally consider the differential energy conservation law.
Equation (\ref{energy-eq}) is not suitable for this purpose,
however, because it is not written explicitly 
in terms of the local energy density.
We hence start from the energy density defined by
\begin{eqnarray}
&&\hspace{-6mm}
{\cal E}(1)\equiv 
\frac{\hbar^{2}}{2m}{\bm \nabla}_{1}'\!\cdot\!{\bm \nabla}_{1}
\langle\psi^{\dagger}_{\cal H}(1')\psi_{\cal H}(1)\rangle\bigr|_{1'=1}
\nonumber \\
&&\hspace{-1mm}+
\frac{1}{2}\!\int\!{\rm d}^{3}r_{1}'\,
V({\bm r}_{1}\!-\!{\bm r}_{1}')
\langle\psi_{\cal H}^{\dagger}(1)\psi_{\cal H}^{\dagger}(1')
\psi_{\cal H}(1')
\psi_{\cal H}(1)\rangle \, ,
\nonumber \\
&&
\label{E-density}
\end{eqnarray}
with $t_{1}'\!=\!t_{1}$ in this expression.
Let us differentiate eq.\ (\ref{E-density}) with respect to time,
eliminate time derivatives of the field operators 
by using eq.\ (\ref{eq-motion}),
and carry out the first-order gradient expansion 
for the interaction term.
These standard procedures\cite{KB62,Zubarev74}
lead to the differential energy conservation law:
\begin{equation}
\frac{\partial{\cal E}({\bm r}t)}{\partial t}+{\bm\nabla}\cdot 
{\bm j}_{\varepsilon}({\bm r}t)=
-{\bm j}({\bm r}t)\cdot{\bm\nabla}U({\bm r}t) \, ,
\label{continuity-E}
\end{equation}
where ${\bm j}_{\varepsilon}$ denotes the energy flux density defined by
\begin{eqnarray}
&&\hspace{-4mm}
{\bm j}_{\varepsilon}({\bm r}t)\equiv
\pm\frac{\hbar^{4}}{4m^{2}}({\bm\nabla}\!-\!
{\bm\nabla}'){\bm\nabla}\!\cdot\!{\bm\nabla}'G_{12}({\bm r}t,{\bm r}'t)
\biggr|_{{\bm r}'={\bm r}} 
\nonumber \\
&&\hspace{10mm}
+\frac{1}{2}\int \!{\rm d}\bar{\bm r}\,V(\bar{\bm r})
\langle \psi_{\cal H}^{\dagger}({\bm r}_{-}t)\,
\hat{\!\bm j}({\bm r}_{+}t)\psi_{\cal H}({\bm r}_{-}t)\rangle
\nonumber \\
&&\hspace{10mm}
-\frac{1}{4}\int\!{\rm d}^{3}\bar{r}\,
\bar{\bm r}\,\frac{\partial V(\bar{\bm r})}{\partial \bar{\bm r}}\cdot
\bigl[\langle \psi_{\cal H}^{\dagger}({\bm r}_{-}t)\,
\hat{\!\bm j}({\bm r}_{+}t)\psi_{\cal H}({\bm r}_{-}t)\rangle
\nonumber \\
&&\hspace{16mm}
+\langle \psi_{\cal H}^{\dagger}({\bm r}_{+}t)\,
\hat{\!\bm j}({\bm r}_{-}t)\psi_{\cal H}({\bm r}_{+}t)\rangle\bigr]\, ,
\label{j_e}
\end{eqnarray}
with
\begin{equation}
\hat{\!\bm j}({\bm r}t)\equiv \frac{\hbar}{2mi}({\bm\nabla}\!-\!{\bm\nabla}')
\psi_{\cal H}^{\dagger}({\bm r}'t)\psi_{\cal H}({\bm r}t)
\bigr|_{{\bm r}'={\bm r}} \, .
\end{equation}
Since ${\cal E}$ and ${\bm j}_{\varepsilon}$ in eq.\ (\ref{continuity-E})
are operated by $\partial_{t}$ and ${\bm \nabla}$, respectively,
eqs.\ (\ref{E-density}) and (\ref{j_e}) should be
evaluated with the local approximation in the first-order gradient expansion.
This can be carried out
with the same procedures as those of deriving 
eq.\ (\ref{tildeTheta^K-p}) and the second term in eq.\ (\ref{tildePi2})
from eqs.\ (\ref{Pi^K}) and (\ref{Pi^V}), respectively.
We finally obtain
\begin{subequations}
\label{Ej_e}
\begin{equation}
{\cal E}=\frac{1}{2}mnv^{2}+\tilde{\cal E} \, ,
\end{equation}
\begin{equation}
{\bm j}_{\varepsilon}=\frac{1}{2}mnv^{2}{\bm v}+\tilde{\cal E}{\bm v}
+\underline{\Pi}{\bm v}+{\bm j}_{Q} \, .
\end{equation}
\end{subequations}
Here tensor $\underline{\Pi}$ is given by eq.\ (\ref{tildePi2}), 
and $\tilde{\cal E}$ and ${\bm j}_{Q}$ are defined by
\begin{subequations}
\label{tildeEj_e}
\begin{eqnarray}
&&\hspace{-15mm}
\tilde{\cal E}({\bm r}t)\equiv
\hbar\int\!\frac{{\rm d}^{3}p\,{\rm d}\varepsilon}{(2\pi\hbar)^{4}}\frac{\bar{p}^{2}}{2m}
A({\bm p}\varepsilon,{\bm r}t)
\phi({\bm p}\varepsilon,{\bm r}t) 
\nonumber \\
&&\hspace{-3mm}
\pm \frac{i\hbar}{2} 
\int \frac{{\rm d}^{3}p\,{\rm d}\varepsilon}{(2\pi\hbar)^{4}}
[\Sigma^{\rm R}({\bm p}\varepsilon,{\bm r}t)
G_{12}({\bm p}\varepsilon,{\bm r}t)
\nonumber \\
&&\hspace{-3mm}
+
\Sigma_{12}({\bm p}\varepsilon,{\bm r}t)
G^{\rm A}({\bm p}\varepsilon,{\bm r}t)]\, ,
\label{tildeE}
\end{eqnarray}
\begin{eqnarray}
&&\hspace{-6mm}
{\bm j}_{Q}({\bm r}t)\equiv
\hbar\int\!\frac{{\rm d}^{3}p\,{\rm d}\varepsilon}
{(2\pi\hbar)^{4}}\frac{\bar{p}^{2}}{2m^{2}}\bar{\bm p}
A({\bm p}\varepsilon,{\bm r}t)
\phi({\bm p}\varepsilon,{\bm r}t)
\nonumber \\
&&\hspace{-1mm}
+\frac{(i\hbar)^{2}}{2}\int 
\frac{{\rm d}^{3}q\,{\rm d}\omega}{(2\pi\hbar)^{4}}
\int \frac{{\rm d}^{3}p\,{\rm d}\varepsilon}{(2\pi\hbar)^{4}}
\int \frac{{\rm d}^{3}p'{\rm d}\varepsilon'}{(2\pi\hbar)^{4}}
\nonumber \\
&&\hspace{-1mm}
\times \frac{1}{m}\!\left(2\bar{\bm p} V_{q}
+{\bm q} \frac{{\bm q}\cdot\bar{\bm p}}{q}
\frac{d V_{q}}{d q}\right)\!
\bigl[{\cal K}_{12,12}({\bm p}\varepsilon,{\bm p}'\varepsilon';{\bm q}\omega,{\bm r}t)
\nonumber \\
&&\hspace{-1mm}
+(2\pi\hbar)^{4}
\delta({\bm q})\delta(\omega)G_{12}({\bm p}\varepsilon,{\bm r}t)
G_{12}({\bm p}'\varepsilon'\!,{\bm r}t)
\bigr]  ,
\end{eqnarray}
\end{subequations}
respectively,
with $\bar{\bm p}\!=\!{\bm p}\!-\!m{\bm v}$.
Use has been made of eq.\ (\ref{Hint-identity}) to derive
the interaction term in eq.\ (\ref{tildeE}).
The quantity $\tilde{\cal E}$ denotes the energy density in the reference
frame moving with the local velocity ${\bm v}$, whereas ${\bm j}_{Q}$
is the heat-flux density.

Let us substitute eq.\ (\ref{Ej_e}) into eq.\ (\ref{continuity-E}) and
transform it with eqs.\ (\ref{continuity}) and (\ref{NS}).
We thereby arrive at an alternative expression of 
the differential energy conservation law as
\begin{equation}
\frac{\partial\tilde{\cal E}}{\partial t}+
{\bm\nabla}\cdot(\tilde{\cal E}{\bm v}+{\bm j}_{Q})
+\sum_{ij}\Pi_{ij}\frac{\partial v_{j}}{\partial r_{i}}=0 \, .
\label{continuity-E2}
\end{equation}
Equations (\ref{continuity}), (\ref{NS}) and (\ref{continuity-E2})
with eqs.\ (\ref{nj-p}), (\ref{tildePi2}) and (\ref{tildeEj_e}) 
completely describe the flows of particle, momentum and energy, respectively.

\section{Calculation of $\Phi$ in equilibrium}
\label{app:Phi}

Using the zero-temperature time-ordered Goldstone technique,\cite{FW}
Carneiro and Pethick\cite{CP75} performed a third-order calculation of 
the functional $\Phi$ for a uniform Fermi system.
They thereby found a singular or on-energy-shell contribution
to $\Phi$; see eq.\ (29) and \S IV of their paper. 
It is this on-energy-shell contribution that brings
the difference between eq.\ (\ref{entropy-density})
and the Carneiro-Pethick expression.
As they mentioned explicitly in \S IV.A,
however, there may be ambiguity in the Goldstone technique on
how to regularize the energy denominators.
We here calculate the same contribution to $\Phi$
with the finite-temperature Matsubara formalism, choosing the chemical potential
as an independent variable instead of the total particle number.
The Matsubara formalism has definite advantages 
over the Goldstone technique that 
(i) it can describe finite temperatures
and (ii) no additional regularization procedure is required.

\begin{figure}[t]
\begin{center}
  \includegraphics[width=0.4\linewidth]{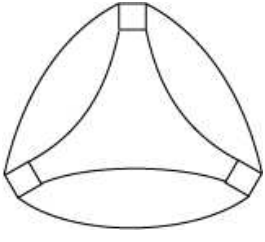}
\end{center}
  \caption{Third-order diagram for $\Phi$.}
  \label{fig:5}
\end{figure}

Let us express the Matsubara
Green's function in the Lehmann representation as
\begin{equation}
G({\bm r}_{1},{\bm r}_{2},z_{n})=\int_{-\infty}^{\infty} 
\frac{{\rm d}\varepsilon}{2\pi}\,\frac{A({\bm r}_{1},{\bm r}_{2},\varepsilon)}
{z_{n}-\varepsilon} \, ,
\label{G-Matsu}
\end{equation}
where $z_{n}\!\equiv \!2n\pi k_{\rm B}T i$ for bosons and
$z_{n}\!\equiv\!(2n\!+\!1)\pi k_{\rm B}T i$ for fermions.
We also introduce the bare vertex:
\begin{eqnarray}
&&\hspace{-5mm}
\langle{\bm r}_{1}'{\bm r}_{2}'|V|{\bm r}_{1}{\bm r}_{2}\rangle
\equiv V({\bm r}_{1}\!-\!{\bm r}_{2})\bigl[\delta({\bm r}_{1}'\!-\!{\bm r}_{1})
\delta({\bm r}_{2}'\!-\!{\bm r}_{2})
\nonumber \\
&&\hspace{21mm} \pm
\delta({\bm r}_{1}'\!-\!{\bm r}_{2})
\delta({\bm r}_{2}'\!-\!{\bm r}_{1})\bigr]\, ,
\label{V-symm}
\end{eqnarray}
which is expressed by a square in the Feynman diagram
following the convention of Abrikosov, Gor'kov and Dzyaloshinski.\cite{AGD}
With eq.\ (\ref{V-symm}), all the third-order contributions to $\Phi$
can be expressed by a single diagram of
Fig.\ \ref{fig:5}.
Using eqs.\ (\ref{G-Matsu}) and (\ref{V-symm}), we can write down
the corresponding analytic expression compactly as
\begin{eqnarray}
&&\hspace{-7mm}
\Phi^{(3)}=\prod_{j=1}^{6}\int_{-\infty}^{\infty}\frac{{\rm d}\varepsilon_{j}}{2\pi}
\, {\cal K}^{(3)}(\varepsilon_{1},\varepsilon_{2},\varepsilon_{3},
\varepsilon_{4},\varepsilon_{5},\varepsilon_{6})
\nonumber \\
&&\hspace{4mm}\times
{\cal I}^{(3)}(\varepsilon_{1},\varepsilon_{2},\varepsilon_{3},
\varepsilon_{4},\varepsilon_{5},\varepsilon_{6}) \, .
\end{eqnarray}
Here ${\cal I}^{(3)}$ is defined by
\begin{eqnarray}
&&\hspace{-7mm}
{\cal I}^{(3)}(\varepsilon_{1},\varepsilon_{2},\varepsilon_{3},
\varepsilon_{4},\varepsilon_{5},\varepsilon_{6})
\nonumber \\
&&\hspace{-7mm}
=\frac{1}{\beta^{4}}\sum_{n_{1}\cdots n_{6}}\frac{
\delta_{n_{1}+n_{2},n_{3}+n_{4}}\delta_{n_{1}+n_{2},n_{5}+n_{6}}}{
(z_{1}\!-\!\varepsilon_{1})(z_{2}\!-\!\varepsilon_{2})
(z_{3}\!-\!\varepsilon_{3})(z_{4}\!-\!\varepsilon_{4})}
\nonumber \\
&&\hspace{-3mm}
\times\frac{1}{
(z_{5}\!-\!\varepsilon_{5})(z_{6}\!-\!\varepsilon_{6})}\, ,
\label{calI^(3)}
\end{eqnarray}
with $\beta\!\equiv\!1/k_{\rm B}T$ and $z_{j}\!\equiv\!z_{n_{j}}$.
The other factor ${\cal K}^{(3)}$ is analytic in $\varepsilon_{j}$ and irrelevant
for the present purpose.
Indeed, it is given explicitly as
\begin{eqnarray}
&&\hspace{-5mm}
{\cal K}^{(3)}(\varepsilon_{1},\varepsilon_{2},\varepsilon_{3},
\varepsilon_{4},\varepsilon_{5},\varepsilon_{6})
\nonumber \\
&&\hspace{-5mm}
=\frac{1}{24}
\int \prod_{j=1}^{6} {\rm d}^{3}r_{j}{\rm d}^{3}r_{j}'\,
\langle{\bm r}_{5}'{\bm r}_{6}'|V|{\bm r}_{5}{\bm r}_{6}\rangle
\langle{\bm r}_{3}'{\bm r}_{4}'|V|{\bm r}_{3}{\bm r}_{4}\rangle
\nonumber \\
&&\hspace{-3mm}
\times
\langle{\bm r}_{1}'{\bm r}_{2}'|V|{\bm r}_{1}{\bm r}_{2}\rangle
A({\bm r}_{1},{\bm r}_{5}',\varepsilon_{5})
A({\bm r}_{5},{\bm r}_{3}',\varepsilon_{3})
A({\bm r}_{3},{\bm r}_{1}',\varepsilon_{1}) 
\nonumber \\
&&\hspace{-3mm}
\times
\bigl[\mp 4A({\bm r}_{2},{\bm r}_{4}',-\varepsilon_{2})
A({\bm r}_{4},{\bm r}_{6}',-\varepsilon_{4})
A({\bm r}_{6},{\bm r}_{2}',-\varepsilon_{6})
\nonumber \\
&&\hspace{1mm}
+A({\bm r}_{2},{\bm r}_{6}',\varepsilon_{6})
A({\bm r}_{6},{\bm r}_{4}',\varepsilon_{4})
A({\bm r}_{4},{\bm r}_{2}',\varepsilon_{2})\bigr]\, ,
\end{eqnarray}
where the first (second) term in the square bracket
corresponds to the ring and particle-hole (particle-particle)
contribution.

The whole issue here is whether the above ${\cal I}^{(3)}$
contains the on-energy-shell contribution.
However, the expression (\ref{calI^(3)}) already tells us the
absence of the on-energy-shell contribution for the Fermi system
considered by Carneiro and Pethick:\cite{CP75} 
because $z_{j}$ is pure imaginary and $\varepsilon_{j}$ is real,
the fraction is regular at any finite temperature,
even for $\varepsilon_{1}\!+\!\varepsilon_{2}\!=\!
\varepsilon_{3}\!+\!\varepsilon_{4}
\!=\!\varepsilon_{5}\!+\!\varepsilon_{6}$.
We shall confirm this fact further below.

We now carry out the summations over $n_{j}$ one by one 
with the standard technique of transforming 
them into contour integrals with\cite{FW}
\begin{equation}
\phi(z)=\frac{1}{{\rm e}^{\beta z}\mp 1} \, .
\end{equation}
First, those over $n_{3}$ and $n_{4}$ in eq.\ (\ref{calI^(3)}) 
yield
\begin{eqnarray}
&&\hspace{-10mm}
\frac{1}{\beta}\sum_{n_{3}n_{4}}\frac{\delta_{n_{1}+n_{2},n_{3}+n_{4}}}
{(z_{3}\!-\!\varepsilon_{3})(z_{4}\!-\!\varepsilon_{4})}
=\frac{\phi_{3}\phi_{4}\!-\! (1\!\pm\! \phi_{3})(1\!\pm\! \phi_{4})}
{z_{1}\!+\! z_{2}\!-\! \varepsilon_{3}\!-\! \varepsilon_{4}}
\nonumber \\
&&\hspace{-10mm}
\equiv {\cal J}^{(2)}(z_{1},z_{2},\varepsilon_{3},\varepsilon_{4}) \, ,
\label{calI^(2a)}
\end{eqnarray}
with $\phi_{j}\!\equiv\! \phi(\varepsilon_{j})$.
In obtaining the result, 
we have used $\phi(z_{1}\!+\!z_{2}\!-\!\varepsilon_{4})
\!=\!\phi(-\varepsilon_{4})\!=\!\mp (1\! \pm\!\phi_{4})$
and $1\!\pm\! \phi_{3}\!\pm\!\phi_{4}\!=\! (1\!\pm \!\phi_{3})(1\!\pm \!\phi_{4})\!-\!
\phi_{3}\phi_{4}$. Note that ${\cal J}^{(2)}$ is regular for 
the imaginary arguments $z_{1}$ and  $z_{2}$.
The summations over $n_{5}$ and $n_{6}$ can be performed
similarly. 

Before proceeding directly to the summation over $n_{2}$ 
in eq.\ (\ref{calI^(3)}), it is useful to consider a couple 
of summations connected with eq.\ (\ref{calI^(2a)}).
The first one ${\cal S}^{(2)}$ is defined and transformed as follows:
\begin{eqnarray}
&&\hspace{-10mm}
{\cal S}^{(2)}(z_{1},\varepsilon_{2},\varepsilon_{3},\varepsilon_{4})
\equiv \frac{1}{\beta}\sum_{n_{2}}\frac{{\cal J}^{(2)}(z_{1},z_{2},\varepsilon_{3},
\varepsilon_{4})}{z_{2}\!-\!
\varepsilon_{2}}
\nonumber \\
&&\hspace{-10mm}
= \pm \frac{ \phi_{2}(1\!\pm \!\phi_{3})(1\!\pm \!\phi_{4})\!\mp \!(1\!\pm \!\phi_{2})
\phi_{3}\phi_{4}}{z_{1}\!+\! \varepsilon_{2}\!-\! \varepsilon_{3}\!-\! \varepsilon_{4}} \, ,
\label{calS^(2)}
\end{eqnarray}
where we have used $1\!\pm\! \phi_{3}\!\pm\!\phi_{4}\!=\!\phi_{3}\phi_{4}[{\rm e}^{\beta(
\varepsilon_{3}+\varepsilon_{4})}\!-\! 1]$ and 
$[{\rm e}^{\beta(\varepsilon_{3}+\varepsilon_{4})}\!-\! 1]
\phi(\varepsilon_{3}\!+\!\varepsilon_{4}\!-\!z_{1})\!=\!\pm 1$.
The second one ${\cal I}^{(2)}$ is given by
\begin{subequations}
\begin{eqnarray}
&&\hspace{-15mm}
{\cal I}^{(2)}(\varepsilon_{1},\varepsilon_{2},\varepsilon_{3},\varepsilon_{4})
\equiv \frac{1}{\beta}\sum_{n_{1}}\frac{{\cal S}^{(2)}(z_{1},\varepsilon_{2},
\varepsilon_{3},\varepsilon_{4})}{z_{1}\!-\!\varepsilon_{1}}
\nonumber \\
&&\hspace{-15mm}
= \frac{(1\!\pm \!\phi_{1})(1\!\pm \!\phi_{2})\phi_{3}\phi_{4}
- \phi_{1}\phi_{2}(1\!\pm \!\phi_{3})(1\!\pm \!\phi_{4})}
{\varepsilon_{1}\!+\! \varepsilon_{2}\!-\! \varepsilon_{3}\!-\! \varepsilon_{4}} \, ,
\label{calI^(2)}
\end{eqnarray}
where use has been made of the identity
$[\phi_{2}(1\!\pm \!\phi_{3})(1\!\pm \!\phi_{4})\!\mp \!(1\!\pm \!\phi_{2})
\phi_{3}\phi_{4}]
\phi(\varepsilon_{3}\!+\! \varepsilon_{4}\!-\! \varepsilon_{2})
\!=\! (1\!\pm\!\phi_{2})\phi_{3}\phi_{4}$.
Equation (\ref{calI^(2)}) is what we encounter in the second-order
calculation for $\Phi$.\cite{Kita99}
Note that ${\cal I}^{(2)}$ is analytic even when 
$\varepsilon_{1}\!+\! \varepsilon_{2}\!=\! \varepsilon_{3}\!+\! \varepsilon_{4}$.
Thus, we may write ${\cal I}^{(2)}$ alternatively as
\begin{eqnarray}
&&\hspace{-14mm}
{\cal I}^{(2)}(\varepsilon_{1},\varepsilon_{2},\varepsilon_{3},\varepsilon_{4})
\nonumber \\
&&\hspace{-14mm}
={\rm P} \frac{(1\!\pm \!\phi_{1})(1\!\pm \!\phi_{2})\phi_{3}\phi_{4}}
{\varepsilon_{1}\!+\! \varepsilon_{2}\!-\! \varepsilon_{3}\!-\! \varepsilon_{4}}
-{\rm P}\frac{\phi_{1}\phi_{2}(1\!\pm \!\phi_{3})(1\!\pm \!\phi_{4})}
{\varepsilon_{1}\!+\! \varepsilon_{2}\!-\! \varepsilon_{3}\!-\! \varepsilon_{4}},
\label{calI^(2)-2}
\end{eqnarray}
\end{subequations}
with P denoting the principal value.

Now, after those over $n_{3}$, $n_{4}$, $n_{5}$ and $n_{6}$ 
given by eq.\ (\ref{calI^(2a)}),
the summation over $n_{2}$ in eq.\ (\ref{calI^(3)}) is performed as follows:
\begin{eqnarray}
&& \hspace{-10mm}
{\cal S}^{(3)}(z_{1},\varepsilon_{2},\varepsilon_{3},\varepsilon_{4},
\varepsilon_{5},\varepsilon_{6})
\nonumber \\
&&\hspace{-10mm}
\equiv\frac{1}{\beta}\sum_{n_{2}}
\frac{{\cal J}^{(2)}(z_{1},z_{2},\varepsilon_{3},\varepsilon_{4})
{\cal J}^{(2)}(z_{1},z_{2},\varepsilon_{5},\varepsilon_{6})}
{z_{2}\!-\! \varepsilon_{2}}
\nonumber \\
&&\hspace{-10mm}
= 
\bigl\{{\cal S}^{(2)}(z_{1},\varepsilon_{2},\varepsilon_{3},\varepsilon_{4})[
\phi_{5}\phi_{6}\!-\!(1\!\pm\!\phi_{5})(1\!\pm\!\phi_{6})]
\nonumber \\
&&\hspace{-6.8mm}
-{\cal S}^{(2)}(z_{1},\varepsilon_{2},\varepsilon_{5},\varepsilon_{6})[
\phi_{3}\phi_{4}\!-\!(1\!\pm\!\phi_{3})(1\!\pm\!\phi_{4})] \bigr\}
\nonumber \\
&&\hspace{-6.8mm}
\times\frac{1}
{\varepsilon_{3}\!+\! \varepsilon_{4}\!-\! \varepsilon_{5}\!-\! \varepsilon_{6}}\, .
\label{calS^(3)}
\end{eqnarray}
This ${\cal S}^{(3)}$ is clearly analytic at
$\varepsilon_{3}\!+\! \varepsilon_{4}\!=\! \varepsilon_{5}\!+\! \varepsilon_{6}$.
Finally, eq.\ (\ref{calI^(3)}) is transformed as
\begin{eqnarray}
&& \hspace{-10mm}
{\cal I}^{(3)}(\varepsilon_{1},\varepsilon_{2},\varepsilon_{3},\varepsilon_{4},
\varepsilon_{5},\varepsilon_{6})
\nonumber \\
&&\hspace{-10mm}
=\frac{1}{\beta}\sum_{n_{1}}
\frac{{\cal S}^{(3)}(z_{1},\varepsilon_{2},\varepsilon_{3},\varepsilon_{4}
,\varepsilon_{5},\varepsilon_{6})}
{z_{1}\!-\! \varepsilon_{1}}
\nonumber \\
&&\hspace{-10mm}
= 
\bigl\{{\cal I}^{(2)}(\varepsilon_{1},\varepsilon_{2},\varepsilon_{3},\varepsilon_{4})[
\phi_{5}\phi_{6}\!-\!(1\!\pm\!\phi_{5})(1\!\pm\!\phi_{6})]
\nonumber \\
&&\hspace{-6.8mm}
-{\cal I}^{(2)}(\varepsilon_{1},\varepsilon_{2},\varepsilon_{5},\varepsilon_{6})[
\phi_{3}\phi_{4}\!-\!(1\!\pm\!\phi_{3})(1\!\pm\!\phi_{4})] \bigr\}
\nonumber \\
&&\hspace{-6.8mm}
\times\frac{1}
{\varepsilon_{3}\!+\! \varepsilon_{4}\!-\! \varepsilon_{5}\!-\! \varepsilon_{6}}\, .
\label{calI^(3)-2}
\end{eqnarray}
Remember that ${\cal I}^{(2)}$ is analytic as eq.\ (\ref{calI^(2)}).
In addition, ${\cal I}^{(3)}$ has no singularity at
$\varepsilon_{3}\!+\! \varepsilon_{4}\!=\! \varepsilon_{5}\!+\! \varepsilon_{6}$.
We hence conclude that ${\cal I}^{(3)}$ is analytic, i.e.,
there is no on-energy-shell term in ${\cal I}^{(3)}$.
This is clearly a general feature of $\Phi$, as may be realized most easily
from its expression with respect to the Matsubara frequency 
such as eq.\ (\ref{calI^(3)})

From eqs.\ (\ref{calS^(2)}) and (\ref{calI^(2)-2}), we obtain
\begin{equation}
\frac{\delta {\cal I}^{(2)}(\varepsilon_{1},\varepsilon_{2},\varepsilon_{3},
\varepsilon_{4})}{\delta \phi_{1}}=\mp 
{\rm Re}{\cal S}^{(2)}(\varepsilon_{1+},\varepsilon_{2},\varepsilon_{3},
\varepsilon_{4}) \, ,
\end{equation}
with $\varepsilon_{1+}\!\equiv\!\varepsilon_{1}\!+\!i0_{+}$.
The same relation holds between ${\cal I}^{(3)}$ and 
${\cal S}^{(3)}$. Using them, we obtain eqs.\ (3)-(5) of ref.\ \onlinecite{Kita99}.
It should be noted that the same expression as eq.\ (5) of ref.\ \onlinecite{Kita99}
had been presented by Fulde and Wagner\cite{FW71} for a Bose system without
any detailed derivations.



\begin{thebibliography}{99}
\bibitem{comment1}The term $-1$ in the integrand is absent
in the Boltzmann $H$-function\cite{CC90} but naturally results
from the high-temperature expansion of the quantum statistical
expression.
\bibitem{comment2}We should point out the difference between the
present paper and ref.\ \onlinecite{Kita99} that the total particle number
(the chemical potential) is chosen as an independent variable 
in the present paper (ref.\ \onlinecite{Kita99}).
\end{thebibliography}
\end{document}


\title{Addenda to ``Entropy in Nonequilibrium Statistical Mechanics''}

\author{Takafumi KITA}

\affiliation{Department of Physics, Hokkaido University, Sapporo 060-0810, Japan}

\date{\today}
\maketitle

In a previous paper I have proposed a principle of maximum entropy 
for nonequilibrium steady states.\cite{Kita06-1} 
However, the argument based on the $H$ theorem in Introduction
is not appropriate, 
because there is net outflow of entropy in nonequilibrium steady states. 
Thus, the principle is to be regarded as a pure conjecture.

It should also be noted that the principle is identical in the dilute classical limit
to the information theory developed 
by Jou and co-workers.\cite{Jou93}
Kim and Hayakawa\cite{Kim03} tested the information theory 
for dilute classical hard-core and Maxwell molecules with heat conduction
by comparing its predictions with those from the steady-state Boltzmann equation.
They found that the two approaches can yield identical results only up to the
first-order deviations from the local equilibrium generally, and
discrepancies emerge at the second order for all the physical quantities except entropy.
This implies that the above principle cannot be valid beyond the first order.
It is possible, however, that it generally holds within this lowest order, as exemplified by
Kim and Hayakawa\cite{Kim03} and also by
ref.\ \onlinecite{Kita06-2} on Rayleigh-B\'enard convection.
Another support to this conjecture is that the nonequilibrium statistical operator
of Zubarev corresponds to an extremum of the information entropy.\cite{Zubarev74}
Note finally that, 
unlike the linear response theory which is an expansion from the global equilibrium,
the first-order expansion from the local equilibrium
can describe a wide range of nonequilibrium nonlinear phenomena,
as may be realized by the fact that the Navier-Stokes equations belong to this 
order of approximation.\cite{CC90,Kita06-2}
We certainly need further investigations to clarify the applicable range
of the conjecture.

\begin{acknowledgments}

I am grateful to H. Hayakawa for critical discussions.

\end{acknowledgments}